\documentclass[twocolumn, numberedappendix, apj, 12pt]{openjournal} % OJAp

\usepackage{newtxtext,newtxmath}
\usepackage[T1]{fontenc}

\usepackage{graphicx}	% Including figure files
\usepackage{amsmath}     % Advanced math
\usepackage{amsfonts}    % Advanced math
\usepackage{amssymb}     % Advanced math
\usepackage{booktabs}
\usepackage{enumitem}
\usepackage[dvipsnames]{xcolor}
\usepackage[colorlinks]{hyperref}
\hypersetup{
colorlinks=true,
urlcolor=[HTML]{8A0087},
filecolor=[HTML]{131877},
citecolor=NavyBlue,
linkcolor=NavyBlue,
}
\usepackage{orcidlink}   % orcidlink
\usepackage{lineno}     % linenumbers
\newcommand{\Mtwooo}{\ensuremath{M_{200\mathrm{c}}}}
\newcommand{\Mwl}{\ensuremath{M_\mathrm{WL}}}
\newcommand{\Ctwooo}{\ensuremath{c_{200\mathrm{c}}}}
\newcommand{\redshift}{\ensuremath{z}}
\newcommand{\clz}{\ensuremath{z_\mathrm{cl}}}
\newcommand{\mass}{\ensuremath{M}}
\newcommand{\dif}{\ensuremath{\mathrm{d}}}
\newcommand{\aperKappa}{\ensuremath{M_{\kappa}}}

\newcommand{\thetas}{\ensuremath{\theta_{\mathrm{s}}}}

\newcommand{\numin}{\ensuremath{\nu_{\mathrm{min}}}}
\newcommand{\thetaout}{\ensuremath{\theta_{\mathrm{out}}}}
\newcommand{\vect}[1]{\boldsymbol{\mathbf{#1}}}
\renewcommand{\andname}{\ignorespaces}

\graphicspath{{./}{figures/}}
\begin{document}

%%%%%%%%%%%%%%%%%%% TITLE PAGE %%%%%%%%%%%%%%%%%%%

\title{Weak-Lensing Shear-Selected Galaxy Clusters from the Hyper Suprime-Cam Subaru Strategic Program: \\I. Cluster Catalog, Selection Function and Mass--Observable Relation\vspace{-4.5em}} 
\shorttitle{Mass--Observable Relation of HSC Shear-selected Clusters}
\shortauthors{Chen et al.}

% The list of authors
\author{Kai-Feng~Chen$^{1, 2, \star}$\orcidlink{0000-0002-3839-0230}}
\author{I-Non~Chiu$^{3}$\orcidlink{0000-0002-5819-6566}}
\author{Masamune~Oguri$^{4, 5}$\orcidlink{0000-0003-3484-399X}}

\author{Yen-Ting~Lin$^{6}$\orcidlink{0000-0001-7146-4687}}
\author{Hironao~Miyatake$^{7,8,9}$\orcidlink{0000-0001-7964-9766}}
\author{Satoshi Miyazaki$^{10}$\orcidlink{0000-0002-1962-904X}}
\author{Surhud~More$^{11,9}$\orcidlink{0000-0002-2986-2371}}
\author{Takashi~Hamana$^{12}$}
\author{Markus~M.~Rau$^{13,14}$\orcidlink{0000-0003-3709-1324}}
\author{Tomomi~Sunayama$^{15,7}$\orcidlink{0009-0004-6387-5784}}
\author{Sunao~Sugiyama$^{16,9}$\orcidlink{0000-0003-1153-6735}}
\author{Masahiro~Takada$^{9}$\orcidlink{0000-0002-5578-6472}}

\affiliation{$^{1}$MIT Kavli Institute, Massachusetts Institute of Technology, Cambridge, MA 02139, USA}
\affiliation{$^{2}$Department of Physics, Massachusetts Institute of Technology, Cambridge, MA 02139, USA}
\affiliation{$^{3}$Department of Physics, National Cheng Kung University, 70101 Tainan, Taiwan}
\affiliation{$^{4}$Center for Frontier Science, Chiba University, 1-33 Yayoi-cho, Inage-ku, Chiba 263-8522, Japan}
\affiliation{$^{5}$Department of Physics, Graduate School of Science, Chiba University, 1-33 Yayoi-Cho, Inage-Ku, Chiba 263-8522, Japan}
\affiliation{$^{6}$Institute of Astronomy and Astrophysics Academia Sinica, Taipei 106216, Taiwan}
\affiliation{$^{7}$Kobayashi-Maskawa Institute for the Origin of Particles and the Universe (KMI), Nagoya University, Nagoya, 464-8602, Japan}
\affiliation{$^{8}$Institute for Advanced Research, Nagoya University, Nagoya 464-8601, Japan}
\affiliation{$^{9}$Kavli Institute for the Physics and Mathematics of the Universe (WPI), The University of Tokyo Institutes for Advanced Study (UTIAS), The University of Tokyo, Chiba 277-8583, Japan}
\affiliation{$^{10}$Subaru Telescope, National Astronomical Observatory of Japan, 650 North Aohoku Place Hilo, HI 96720, USA}
\affiliation{$^{11}$The Inter-University Centre for Astronomy and Astrophysics, Post bag 4, Ganeshkhind, Pune 411007, India}
\affiliation{$^{12}$National Astronomical Observatory of Japan, National Institutes of Natural Sciences, Mitaka, Tokyo 181-8588, Japan}
\affiliation{$^{13}$High Energy Physics Division, Argonne National Laboratory, Lemont, IL 60439, USA}
\affiliation{$^{14}$McWilliams Center for Cosmology, Department of Physics, Carnegie Mellon University, 5000 Forbes Avenue, Pittsburgh, PA 15213, USA}
\affiliation{$^{15}$Department of Astronomy and Steward Observatory, University of Arizona, 933 North Cherry Avenue, Tucson, AZ 85719, USA}
\affiliation{$^{16}$Center for Particle Cosmology, Department of Physics and Astronomy, University of Pennsylvania, Philadelphia, PA 19104, USA}
\thanks{$^{\star}$E-mail:kfchen@mit.edu}

%%%%%%%%%%%%%%%%%%%%%%%%%%%%%%%%%%%%%%%%%%%%%%%%%%
%
% Abstract
%
%%%%%%%%%%%%%%%%%%%%%%%%%%%%%%%%%%%%%%%%%%%%%%%%%%

\begin{abstract}
We present the first step toward deriving cosmological constraints through the abundances of galaxy clusters selected in a $510\,\mathrm{deg}^2$ weak-lensing aperture mass map, constructed with the Year-Three shear catalog from the Hyper Suprime-Cam Subaru Strategic Program. We adopt a conservative source galaxy selection to construct a sample of $129$ weak-lensing peaks with a signal-to-noise ratio above $4.7$. We use semi-analytical injection simulations to derive the selection function and the mass--observable relation of our sample. These results take into account complicated uncertainties associated with weak-lensing measurements, such as the non-uniform survey depth and the complex survey geometry, projection effects from uncorrelated large-scale structures, and the intrinsic alignment of source galaxies. We also propose a novel modeling framework to make parts of the mass--observable relation insensitive to assumed cosmological parameters. Such a framework not only offers a great computational advantage to cosmological studies, but can also benefit future astrophysical studies using shear-selected clusters. Our results are an important step toward utilizing these cluster samples that are constructed nearly independent of any baryonic assumptions in upcoming deep-and-wide lensing surveys from the Vera Rubin Observatory, Euclid, and the Nancy Grace Roman Space Telescope.
\end{abstract}

%%%%%%%%%%%%%%%%%%%%%%%%%%%%%%%%%%%%%%%%%%%%%%%%%%
%
% Introduction
%
%%%%%%%%%%%%%%%%%%%%%%%%%%%%%%%%%%%%%%%%%%%%%%%%%%

\section{Introduction} \label{sec:intro}

Observations of large-scale structures (LSS) probe the cosmic expansion history and provide stringent constraints on the matter-energy content of the universe. Over the last few decades, the standard $\Lambda$ Cold Dark Matter ($\Lambda$CDM) paradigm has achieved great phenomenological success in describing these observations \citep[e.g.,][]{Frieman:2008, Weinberg:2013}. However, the increasingly precise cosmological measurements have raised the question of whether observations from different tracers across a wide range of cosmic time can be consistently explained by the same set of parameters in the $\Lambda$CDM model \citep[see][for a recent summary]{S8review:2022}. While these discrepancies among observations can be an exciting indicator for new physics beyond $\Lambda$CDM, they are equally likely to be caused by unaccounted-for systematic effects or inadequate uncertainty modeling. Accurately characterizing observational data with a comprehensive model of known physical and systematic effects is therefore an important task for cosmologists in this decade. 

An example of these tensions is in the measurements of $S_8\coloneqq\sigma_8\sqrt{\Omega_\mathrm{m}/0.3}$, the present-day amplitude of the matter fluctuations $\sigma_8$ scaled by the square root of the total matter density $\Omega_\mathrm{m}$. Early-universe observations from the cosmic microwave background \citep[CMB, ][]{Planck18, ACT+WMAP:2020} suggest a higher $S_8$ than those inferred from the late universe with cosmic shear \citep[e.g.][]{KiDS1000CosmiShear:2021, DESY3CosmicShear:2022a, DESY3CosmicShear:2022b, HSCY32ptCF, HSCY3CosmicShearPS}, galaxy clustering and galaxy-galaxy lensing \citep{[e.g.][]DESY3_redMaGiC2x2:2022, DESY3_MAGLIM2x2:2022}, cluster abundance \citep[e.g.][]{PlanckSZ, Bocquet:2019, Chiu:2023}, and a combination of these probes \citep[e.g.][]{DESY1_6x2+N:2021, KiDS1000_3x2pt:2021, DESY3_3x2:2022, HSCY3_3x2pt-a:2023, HSCY3_3x2pt-b:2023}.

In particular, observations utilizing the effect of weak gravitational lensing \citep[hereinafter weak lensing; see][for a comprehensive review]{Mandelbaum:2018} offer some of the strongest constraints on $S_8$ from the late universe. These observations measure small but coherent shape distortions, often referred to as shears, of distant galaxies (source) by intervening foreground structures (lens), giving us a direct probe of matter density fluctuations. The first cosmological application of weak lensing comes from the detection of cosmic shear variance \citep{Bacon:2000, Kaiser:2000, VanWaerbeke:2000, Wittman:2000}, and subsequent cosmological constraints from cosmic shear power spectra and correlation functions have shown the power of weak lensing with just the Gaussian summary statistics \citep{Kilbinger:2015}. Meanwhile, it is of increasing interest to extract additional cosmological information from higher-order statistics in weak-lensing observations. Examples of these summary statistics include number counts of peaks \citep{Dietrich:2010, Hamana:2015, Liu_X:2015, Liu_J:2015, Kacprzak:2016, Shan:2018, Martinet:2018, Harnois-Deraps:2021, Zurcher:2022, Liu:2023, Marques:2023} or higher moments from weak-lensing mass maps \citep{VanWaerbek:2013, Petri:2015, Peel:2018, Chang:2018, Gatti:2020, Gatti:2022, Anbajagane:2023}, three-point correlation functions or bispectra \citep{Takada:2003, Takada:2004, Dodelson:2005, Vafaei:2010, Berge:2010, Semboloni:2011, Fu:2014}, Minkowski functionals \citep{Kratochvil:2012, Petri:2013, Vicinanza:2019, Parroni:2020}, density split statistics \citep{Gruen:2018, Friedrich:2018}, and direct field-level inference \citep{Jeffrey:2020, Fluri:2022}. Obtaining accurate and competitive cosmological constraints from these higher-order statistics will serve as an important consistency test among weak-lensing probes and help break parameter degeneracies.

In a series of two papers \citep[this work and][]{PartII}, we focus on constraining cosmology with high signal-to-noise ratio peaks detected on weak-lensing aperture mass maps \citep{Schneider1996, S19A_S2C}. These peaks are associated with massive dark matter halos where clusters of galaxies reside. The number counts of galaxy clusters, modeled by the halo mass function, are sensitive to both the geometry and the structure formation history in our universe \citep[e.g.,][]{Allen:2011}. Tight cosmological constraints have been obtained with cluster samples selected with X-ray signals \citep{Vikhlinin:2009, Mantz:2010, Mantz:2014, Chiu:2023, eROSITA_CC:2024} from the thermal bremsstrahlung emission, at millimeter wavelengths \citep{PlanckSZ, deHaan:2016, Bocquet:2019} by the thermal Sunyaev–Zel’dovich effect \citep{tSZ}, and in the optical \citep{Rozo:2010, DESY1_CC:2020, DESY1_CC+SPT:2021, AMICO_CC1, AMICO_CC2, Sunayama:2023} via overdensities of galaxies. However, all of these samples rely on baryonic tracers of clusters and require complex astrophysical modeling to account for the selection effect and the mass--observable relation. Incorrect or insufficient modeling of these baryonic effects could lead to biases in the cosmological constraints \citep{Salvati:2020, Grandis:2021}. Cluster samples constructed from weak-lensing maps \citep{Wittman:2001, Schirmer:2007, Miyazaki2007, Shan:2012, Miyazaki:2015, S16A_S2C, S19A_S2C}, on the other hand, allow us to determine the selection effect based purely on the theory of gravity, offering a direct link between the weak-lensing observable and the underlying halo mass. Shear-selected clusters, therefore, provide a powerful cosmology probe that complements both the cosmic shear power spectrum and traditional cluster cosmology.

Despite the direct link between the weak-lensing observable and the underlying projected density field, it is still essential to construct a comprehensive modeling framework that takes into account all physical and systematic effects associated with weak-lensing observations. In the past, weak-lensing peak counts were often modeled analytically by using halo models on a Gaussian random field \citep{Fan:2010, Shan:2018} or semi-analytically by injecting synthetic halo profiles into N-body simulations \citep{Lin-CA:2015}. Important systematic effects such as dilution by cluster member galaxies \citep{HSC-WK_Plank-SZ, Y1_Cluster_WL, S19A_S2C}, intrinsic alignment of source galaxies \citep{Kacprzak:2016, Harnois-Deraps:2021, Zhang:2022}, and other baryon physics \citep{Osato:2015, Weiss:2019, Coulton:2020, Osato:2021, Lee:2023, Broxterman:2024} are hard to model and can only be accounted for using nuisance parameters with strong priors. On the other hand, ray-tracing simulations have also been carried out to study weak-lensing peak counts \citep{Dietrich:2010, Liu_J:2015, Kacprzak:2016}. However, these simulations are generally expensive to compute and can only be carried out on a limited grid of cosmological parameters. 

In this work, we introduce a novel semi-analytical framework to model weak-lensing peak counts. Instead of painting halos into N-body simulations, we inject halos into the observed weak-lensing mass maps. This is achieved by adding the lensing signals from synthetic halos onto the shear of real source galaxies in the shape catalog. As the shape of these source galaxies already contains realistic observational noise and real-world weak-lensing systematic uncertainties, this allows us to accurately characterize the measurement uncertainties on the weak-lensing observable. Meanwhile, by performing injection simulations across the survey footprint, we also take into account non-uniform imaging depth and the complex geometry of the survey due to bright star masks and artifacts. Moreover, we mitigate the contamination from cluster member galaxies by employing a stringent source selection \citep{S19A_S2C} while still maintaining a high source density. This is only possible thanks to the incredible imaging depth provided by the Subaru Hyper Suprime-Cam (HSC) survey \citep{HSC}. While these injection simulations are still expensive to perform, in this work we also introduce a novel parametrization method and choose halo properties that will make the derived selection function and scaling relation independent of the underlying cosmology. The framework and results developed in this paper will be incorporated into \citet{PartII} to obtain cosmological constraints. 

This paper is organized as follows. In Sec.~\ref{sec:data}, we introduce the weak lensing data, the construction of aperture mass maps, and the cluster catalog. Our modeling framework and the details of the injection simulations are discussed in Sec.~\ref{sec:method}. Results of the injection simulations, and the derived selection function and scaling relation are presented in Sec.~\ref{sec:result}. In particular, important validation tests of our modeling framework are shown in Sec.~\ref{subsec:validation}. Conclusions are given in Sec.~\ref{sec:conclusion}.

%%%%%%%%%%%%%%%%%%%%%%%%%%%%%%%%%%%%%%%%%%%%%%%%%%
%
% Data
%
%%%%%%%%%%%%%%%%%%%%%%%%%%%%%%%%%%%%%%%%%%%%%%%%%%

\section{Data} \label{sec:data}
\subsection{Weak-lensing shape catalog} \label{subsec:shear}

The shape catalog from the HSC-SSP S19A internal data release \citep[hereinafter the HSC-Y3 data]{S19A_shear} contains roughly 36 million galaxies across six fields (XMM, GAMA09H, WIDE12H, GAMA15H, VVDS, HECTOMAP)\footnote{The complete HSC survey consists of only three distinct patches. However, as of year three, the survey footprint is still fragmented.} with an averaged $19.9~\mathrm{arcmin}^{-2}$ effective galaxy number density. The galaxy shapes are measured with the re-Gaussianization method \citep{reGauss} and the shear estimation bias is calibrated using image simulations \citep{shape_bias}. {The mean shear values, the stacked cross shear signals around galaxies, and the star-galaxy cross-correlation functions of the Y3 catalog have been demonstrated to be mostly consistent with zero \citep{S19A_shear}, enabling accurate weak-lensing measurements.}

\begin{figure*}[bth]
    \centering  
    \epsscale{1.175}
    \plotone{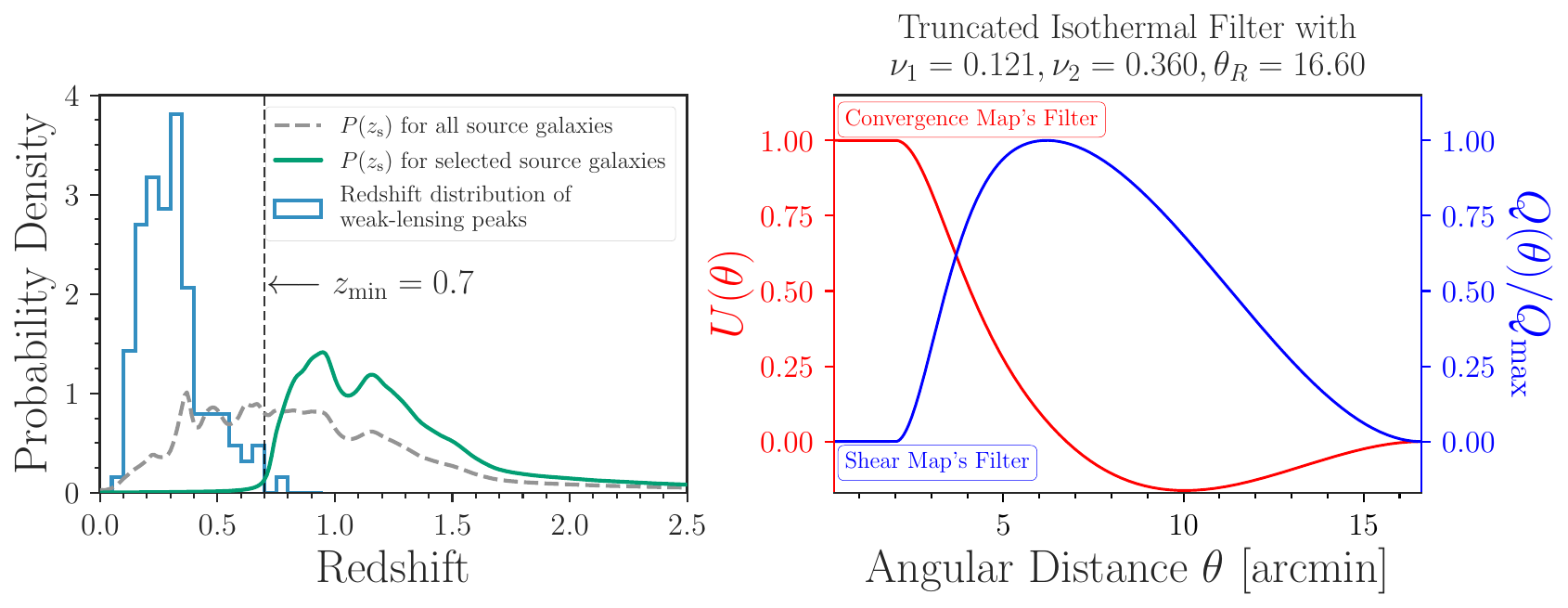}
    \caption{\textit{Left}: probability density function of source galaxies' photometric redshift in the HSC-Y3 shape catalog before (dashed grey) and after (solid green) the selection criterion given in Equation \eqref{eq:pdz_select}. The redshift distribution of weak-lensing peaks with signal-to-noise ratio larger than 4.7 is shown in the blue histogram. The redshift of the weak-lensing peaks is obtained from cross-matching with optical cluster catalogs as described in Sec.~\ref{subsec:cl_sample}. \textit{Right}: aperture mass filters for convergence (red) and tangential shear (blue) adopted in this work. \label{fig:pdz_filter}}
\end{figure*}

To further improve the accuracy of the weak-lensing observable associated with galaxy clusters, we employ a conservative selection of source galaxies depending on their photometric redshift. It is known that cluster member galaxies dilute weak-lensing signals as their shapes are not distorted by the cluster itself, thus reducing the average ellipticity around the galaxy cluster \citep{HSC-WK_Plank-SZ, Y1_Cluster_WL, S19A_S2C}. \citet{S19A_S2C} showed that a redshift cut on the source sample can effectively mitigate the dilution effect for clusters selected in the weak-lensing mass maps. Following their work, we therefore require all source galaxies in our sample to have
\begin{equation}\label{eq:pdz_select}
    \int_{0.7}^{\infty} P(z)\,\mathrm{d}z \geq 0.95.
\end{equation}
Here, the probability density function $P(z)$ for each source galaxy is obtained with an empirical fitting method from the Direct Empirical Photometric code  \citep[\texttt{DEmP};][]{DEMP, HSC_PDZ1, HSC_PDZ2}. With this conservative cut, the total number of source galaxies reduces to roughly 16.9 million galaxies with an averaged $9.4\,\mathrm{arcmin}^{-2}$ effective galaxy number density. The combined probability density function of source galaxies' redshift before and after this selection is shown as the dashed gray and solid green curve in the left panel of \autoref{fig:pdz_filter}.

\begin{figure*}[tbh]
    \centering
    \epsscale{1.175}
    \plotone{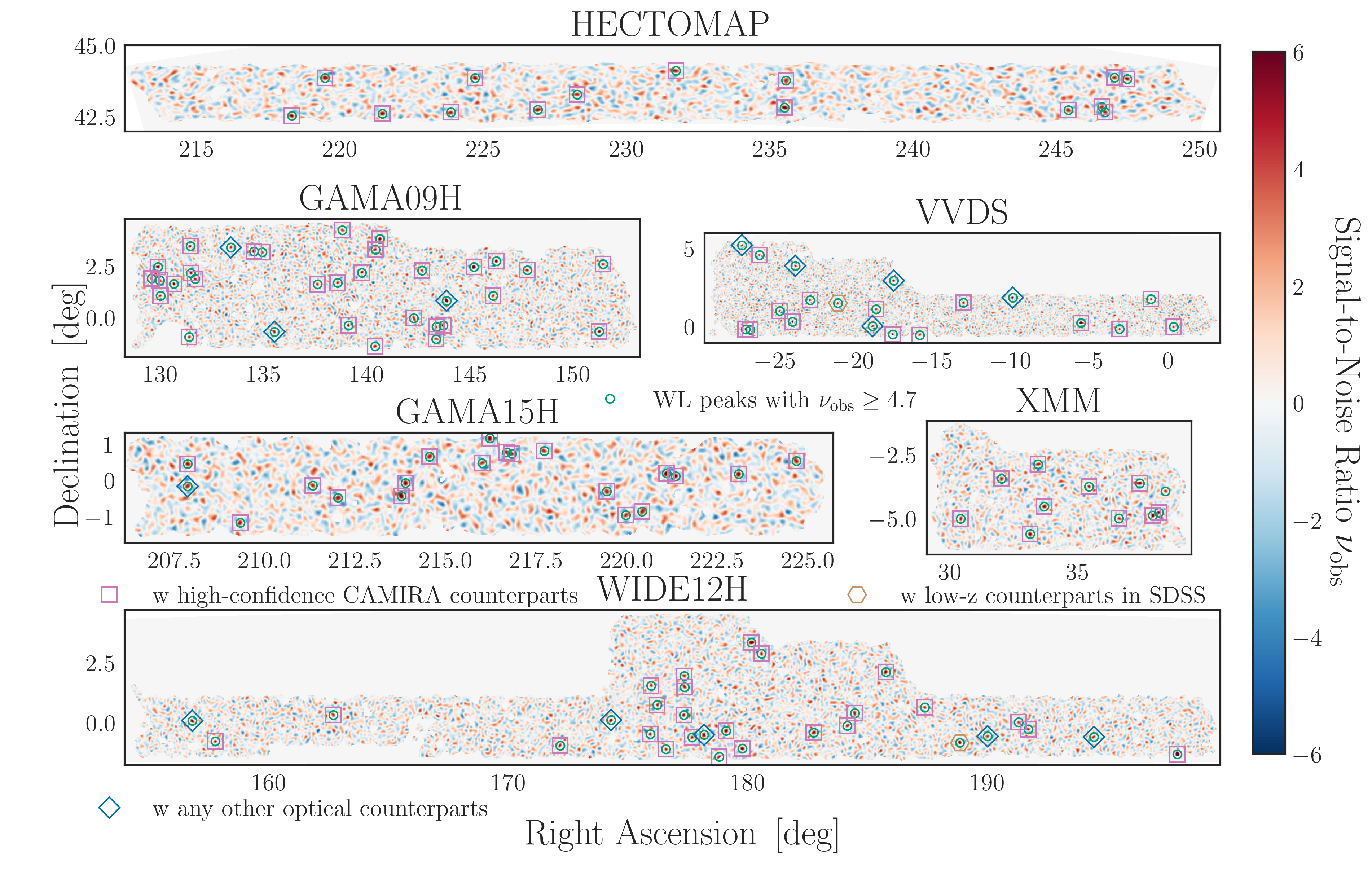}
    \caption{Mass maps derived from the HSC-Y3 data with the truncated isothermal filter given in \autoref{fig:pdz_filter}. Weak-lensing peaks with a signal-to-noise ratio larger than 4.7 are circled in green. These peaks are matched to optical clusters in the following order: CAMIRA cluster with a high $P(N_\mathrm{mem}, \redshift | \nu)$ (purple square); low-redshift clusters found in the SDSS (brown hexagon); any other optical clusters with richness $N_\mathrm{mem} \geq 20$ (blue diamond). \label{fig:mass_map}}
\end{figure*}

\subsection{Construction of Mass Maps} \label{subsec:mass_maps}
To identify signals from galaxy clusters in our weak-lensing shape catalog, we create aperture mass maps $\aperKappa(\vect{\theta})$ \citep{Schneider1996} defined as 
\begin{equation}
    \aperKappa(\vect{\theta})=\int \kappa(\vect{\theta^{\prime}}) U(|\vect{\theta^{\prime}} - \vect{\theta}|)\,\dif\vect{\theta^{\prime}} ,
\end{equation}
where $\kappa(\vect{\theta})$ is a map of the weak-lensing convergence and $U(\theta)$ is a filter that is chosen to maximize the signals from galaxy clusters. We require the filter to be compensated, i.e.,
\begin{equation} \label{eq:U_normalized}
    \int U(\theta)\theta \,\dif\theta = 0, 
\end{equation}
so that the aperture mass maps can be equivalently expressed in terms of the tangential shear maps \citep{Kaiser_Squires}
\begin{equation}
    \begin{aligned}
            \aperKappa(\vect{\theta}) &=\int \kappa(\vect{\theta^{\prime}}) U(|\vect{\theta^{\prime}} - \vect{\theta}|)\,\dif\vect{\theta^{\prime}} \\
            &= \int \gamma_{+}(\vect{\theta^{\prime}}; \theta) Q(|\vect{\theta^{\prime}} - \vect{\theta}|)\,\dif\vect{\theta^{\prime}},
    \end{aligned}
\end{equation}
where $\gamma_+(\vect{\theta^\prime}; \vect{\theta})$ is the tangential shear at position $\vect{\theta^\prime}$ with respect to $\vect{\theta}$ and the filter $Q(\theta)$ is related to $U(\theta)$ as \citep{Kaiser1994}
\begin{equation}
    Q(\theta) = \frac{2}{\theta^2}\int_0^{\theta}  U(\theta^\prime)\theta^\prime\,\dif\theta^\prime - U(\theta).
\end{equation}
In this work, we adopt a truncated isothermal profile \citep{Schneider1996}
%\begin{widetext}
\begin{equation}
\begin{aligned}
    &U(\theta) = \\
    &\begin{cases}
    1 & \left(\theta \leq x_1 \theta_R\right) \\ 
    \displaystyle\frac{1}{1-c}\left(\frac{x_1 \theta_R}{\sqrt{\left(\theta-x_1 \theta_R\right)^2+\left(x_1 \theta_R\right)^2}}-c\right) & \left(x_1 \theta_R \leq \theta \leq x_2 \theta_R\right) \\ 
    \displaystyle\frac{b}{\theta_R^3}\left(\theta_R-\theta\right)^2\left(\theta-\alpha \theta_R\right) & \left(x_2 \theta_R \leq \theta \leq \theta_R\right) \\ 0 & \left(\theta_R \leq \theta\right)\end{cases}
\end{aligned}
\end{equation}
%\end{widetext}
where the constants $\alpha, b, c$  are chosen so that the filter $U$ and its first derivative are both continuous at $\theta = x_2\theta_R$ and the normalization condition in Equation \eqref{eq:U_normalized} is satisfied. The overall shape of such a filter is designed to roughly match the surface density distribution of a galaxy cluster. 
Meanwhile, a constant filter for the convergence at $\theta \leq x_1\theta_R$ makes us insensitive to the potential non-linear and baryonic systematic uncertainties in the innermost region of a cluster, and a smoothly truncated filter reduces possible discontinuities in our aperture mass maps from the complex survey geometry. Following \cite{S19A_S2C}, we adopt $x_1 = 0.121$, $x_2 = 0.36$, and $\theta_R = 16.6\,\mathrm{arcmin}$, which has been shown to be the most effective in mitigating the dilution effect from member galaxies and in suppressing the noise from uncorrelated large-scale structures. The resulting filter $U(\theta)$ and $Q(\theta)$ is shown in the right panel of \autoref{fig:pdz_filter}. 

Meanwhile, we define the noise of the aperture mass map at each location $\vect{\theta}$ to be the standard deviation of $\aperKappa(\vect{\theta}; \phi_1, \phi_2, \ldots, \phi_n)$ where $\phi_1, \phi_2, \ldots, \phi_n$ are random variables drawn from a uniform distribution between $[0, 2\pi)$ denoting the position angle of source galaxies. We thus have
\begin{equation}
    \begin{aligned}
\sigma^2\left(\vect{\theta}_0\right)= & \frac{1}{2 \pi} \int_0^{2 \pi} d \phi_1 \cdots d \phi_n \aperKappa^2(\vect{\theta}; \phi_1, \ldots, \phi_n) \\
& -\frac{1}{2 \pi}\left[\int_0^{2 \pi} d \phi_1 \cdots d \phi_n \aperKappa(\vect{\theta}; \phi_1, \ldots, \phi_n)\right]^2.
\end{aligned}
\end{equation}
In practice, we draw $500$ realizations of $\aperKappa(\vect{\theta}; \phi_1, \allowbreak \phi_2, \ldots, \phi_n)$ by randomly rotating galaxies and take the standard deviation of them as the noise. The signal-to-noise ratio map $\nu(\vect{\theta})$ is then defined as 
\begin{equation}
    \nu(\vect{\theta}) = \frac{\aperKappa(\vect{\theta})}{\sigma(\vect{\theta})}
\end{equation}
Our weak-lensing maps are sampled on a two-dimensional rectangular grid in the $\vect{\theta}$-coordinate with a pixel size of $0.25$ arcmin. \autoref{fig:mass_map} shows six signal-to-noise maps derived from the truncated isothermal filter. For detailed procedure of calculating these weak-lensing maps, we refer the reader to Sec.\,3.4 of \citet{S19A_S2C}.

\subsection{Cluster catalogs} \label{subsec:cl_sample}
From the weak-lensing maps, we select peaks given a signal-to-noise ratio threshold $\numin$. Peaks are identified as pixels that are higher in value than all eight pixels around them. We discard peaks if they are within a $5$ arcmin radius of any other stronger peaks to avoid double counting a single structure. We also discard peaks if they are situated on the boundaries of the map. To check that the peaks we detected on the aperture mass maps are indeed associated with massive clusters, we cross-match the weak-lensing peaks with various optical cluster catalogs, with a primary focus on the CAMIRA cluster sample \citep{CAMIRA_S16A} derived from the HSC S21A photometric data. We also check the cross-match with the redMaPPer \citep{redMaPPer} and the WHL \citep{WHL12, WHL15} cluster samples constructed from the overlapping Sloan Digital Sky Survey Data Release 8 \citep[SDSS, ][]{SDSS_DR8}, mainly to inspect low redshift clusters at $z<0.1$, a regime where the CAMIRA sample does not cover.

For each weak-lensing peaks, we search for optical clusters within a $10$ arcmin radius and keep clusters that are within $1.5\,h^{-1}\mathrm{Mpc}$ using the redshift of the matched optical clusters. As each weak-lensing peak could be matched to multiple optical clusters, we develop the following framework to determine the best-matched optical counterpart.  First, we search for potential counterparts within the CAMIRA catalog using the above-mentioned criteria based on both angular and physical distances. For each of the cross-matched CAMIRA candidate of the observed weak-lensing peak with a signal-to-noise ratio $\nu$, we adopt the richness-to-mass relation from \citet{Murata2019} to estimate the probability of observing the counterpart with a richness of $N_\mathrm{mem}$ at redshift $z_\mathrm{cl}$, denoting as $P(N_\mathrm{mem}, z_\mathrm{cl} | \nu)$ (See Appendix~\ref{appendix:counterpart} for more details). {Here, the richness parameter $N_\mathrm{mem}$ is an estimate of the number of cluster member galaxies and is derived from a weighted sum of background-subtracted red galaxy number density around the brightest cluster galaxy \citep{CAMIRA, CAMIRA_S16A}.} We define a cross-matched CAMIRA candidate as a \textit{high-confidence counterpart} if the richness and redshift of that cluster fall within the $3\sigma$ confidence interval around the peak of the distribution $P(N_\mathrm{mem}, z_\mathrm{cl} | \nu)$. The best-matched optical counterpart is chosen as the most likely among all the high-confidence counterparts. If a peak does not admit a high-confidence counterpart, we then search for the nearest \textit{low-$\redshift$ counterpart} in the redMaPPer and WHL catalogs that has $\redshift \leq 0.2$ and $N_\mathrm{mem} \geq 20$. Lastly, a peak is said to have \textit{any optical counterpart} if it is matched to any optical clusters with $N_\mathrm{mem} \geq 15$ in either CAMIRA, redMaPPer or WHL cluster catalogs \footnote{{We note that while the definition of richness is different between the three optical catalogs, here we adopt a common selection threshold for simplicity. For the CAMIRA catalog, $N_\mathrm{mem} \geq 15$ roughly corresponds to clusters with $M_\mathrm{200m} \geq 10^{14}~h^{-1}M_\odot$ \citep{CAMIRA_S16A}. }}.

\begin{figure}[tbh]
    \centering
    \epsscale{1.175}
    \plotone{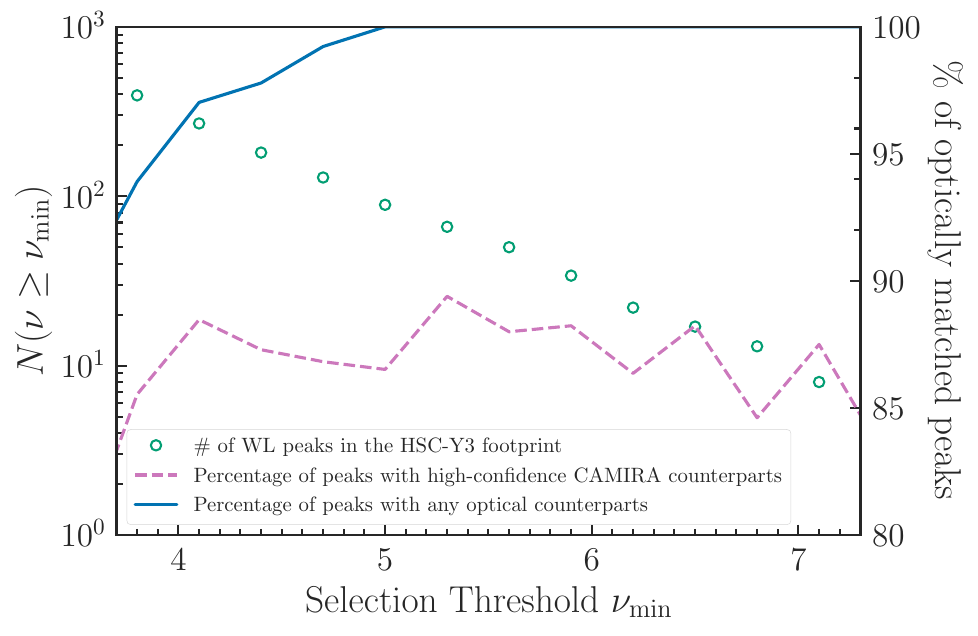}
    \caption{Number of peaks as a function of the selection threshold $\numin$. The percentage of peaks with (high-confidence) optical counterparts is shown in the (dashed purple) solid blue curve.\label{fig:counterpart}}
\end{figure}

\autoref{fig:mass_map} shows an example of a weak-lensing peak sample constructed with $\numin = 4.7$. Under this signal-to-noise ratio threshold, there are in total $129$ peaks identified in the entire HSC-Y3 footprint. Among which, $112$ are found to have high-confidence optical counterparts within the CAMIRA catalog; $2$ peaks do not have high-confidence CAMIRA matches but admit low-$\redshift$ counterparts in SDSS, and $12$ are matched to optical counterparts with $N_\mathrm{mem}\geq 15$ and are within $1.5~h^{-1}\mathrm{Mpc}$ to the peak. While there are $3$ peaks that do not have any optical counterpart under the distance and richness criteria given above, two of them (located at $12\mathrm{h}39\mathrm{m}47.0645\mathrm{s}, -00\mathrm{d}33\mathrm{m}52.8076\mathrm{s}$ and $22\mathrm{h}24\mathrm{m}51.0696\mathrm{s} +03\mathrm{d}52\mathrm{m}47.9063\mathrm{s}$) are surrounded by multiple lower-richness CAMIRA clusters and the remaining one 
($02\mathrm{h}33\mathrm{m}40.6278\mathrm{s} -03\mathrm{d}56\mathrm{m}00.1856\mathrm{s}$) is near optical clusters identified by the Canada-France-Hawaii Telescope at $z\sim0.5$ \citep{CFHT:2011, CFHT:2018}. Therefore, these peaks are likely to be the results of chance projection of multiple halos along the line-of-sight. Such a scenario is taken into account in our modeling framework given in Sec.~\ref{sec:method}. A list of weak-lensing peaks in the HSC-Y3 footprint with $\nu \geq 4.7$ and their best-mateched optical counterpart is given in \autoref{tab:peak_list}. 

The redshift distribution of the best-matched optical counterpart for weak-lensing peaks with $\nu \geq 4.7$ is given as the blue histogram in the left panel of \autoref{fig:pdz_filter}. We see that all but one weak-lensing peaks are located at $z\leq 0.7$. As we only utilize source galaxies with significant probability of being above redshift $0.7$, combining with the fact that we exclude signal from the innermost region of the halo, the weak-lensing observable we obtained is safe from the dilution of cluster member galaxies. 

\autoref{fig:counterpart} shows the number of peaks as a function of the selection threshold $\numin$ and the percentage of peaks with (high confidence) optical counterparts. We observe that $96\%\, (100\%)$ of peaks with $\nu \geq 4.0\,(5.0)$ admit optical counterpart, indicating these weak-lensing peaks to be an effective way to select galaxy clusters without any baryonic assumptions. {Following \citet{S16A_S2C} and \citet{S19A_S2C}, in this work, we focus on peaks with $\nu \geq 4.7$. In Part II of our analysis \citep{PartII}, we will utilize peak samples constructed with different selection threshold $\numin$ to test the robustness of our cosmological constraints. }

We note that while the optical counterparts identified in this section can provide extra information such as redshift and optical richness for the weak-lensing peaks, in this work we merely utilize them to better understand our samples. For our cosmological analysis \citep{PartII}, we will only rely on the weak-lensing observable and do not employ these optical information to derive cosmological constraints. 

\begin{deluxetable*}{ccccccccccc}

\tablecaption{List of weak-lensing peaks in the HSC-Y3 footprint with $\nu \geq 4.7$ and their best-matched optical counterpart\label{tab:peak_list}}

%\tablenum{1}

\tablehead{\colhead{Peak ID} & \colhead{RA} & \colhead{Dec} & \colhead{Peak $\mathcal{S}/\mathcal{N}$} & \colhead{$\textrm{RA}_\mathrm{cl}$} & \colhead{$\textrm{Dec}_\mathrm{cl}$} & \colhead{Redshift} & \colhead{Richness} & \colhead{Separation} & \colhead{Optical Catalog} \\ 
\colhead{} & \colhead{(deg)} & \colhead{(deg)} & \colhead{} & \colhead{(deg)} & \colhead{(deg)} & \colhead{} & \colhead{} & \colhead{($h^{-1}\mathrm{Mpc}$)} & \colhead{}  } 

%% All data must appear between the \startdata and \enddata commands
\startdata
  0 & 197.865421 & -1.320876 & 10.38 & 197.886195 & -1.332920 &  0.18 &166.43 & 0.18 & CAMIRA    \\
  1 & 143.800130 &  +0.813507 & 10.18 & 143.801208 &  +0.825611 &  0.35 & 68.73 & 0.15 & CAMIRA    \\
  2 & 354.413644 &  +0.252496 &  9.41 & 354.415554 &  +0.271375 &  0.25 &111.13 & 0.19 & CAMIRA    \\
  3 & 145.118208 &  +2.452905 &  8.48 & 145.102438 &  +2.477641 &  0.16 & 65.08 & 0.20 & CAMIRA    \\
  4 & 180.079850 &  +3.345811 &  8.05 & 180.105635 &  +3.347094 &  0.14 &114.63 & 0.16 & CAMIRA    \\
  5 & 140.561555 &  +3.804756 &  7.61 & 140.545649 &  +3.778190 &  0.25 & 69.27 & 0.30 & CAMIRA    \\
  6 & 130.592751 &  +1.634875 &  7.34 & 130.591163 &  +1.640656 &  0.42 & 77.22 & 0.08 & CAMIRA    \\
  7 &  37.928678 & -4.887543 &  7.22 &  37.921574 & -4.882613 &  0.19 &115.79 & 0.07 & CAMIRA    \\
  8 & 336.039295 &  +0.323747 &  7.05 & 336.040891 &  +0.325988 &  0.18 & 50.09 & 0.02 & CAMIRA    \\
  9 & 146.207093 &  +2.726571 &  7.02 & 146.196298 &  +2.770387 &  0.22 & 84.54 & 0.40 & CAMIRA    \\
 10 & 188.778711 & -0.827654 &  7.01 & 188.778610 & -0.859100 &  0.11 & 28.23 & 0.16 & WHL       \\
 11 & 221.087980 &  +0.187586 &  6.99 & 221.097548 &  +0.214565 &  0.26 & 36.61 & 0.29 & CAMIRA    \\
 12 & 143.632430 & -0.379970 &  6.87 & 143.630004 & -0.385442 &  0.34 & 46.96 & 0.07 & CAMIRA    \\
 13 & 151.196876 & -0.685035 &  6.76 & 151.215018 & -0.662458 &  0.18 & 42.59 & 0.22 & CAMIRA    \\
 14 & 179.027305 & -0.338319 &  6.70 & 179.045036 & -0.350253 &  0.25 & 83.20 & 0.21 & CAMIRA    \\
 15 & 134.467340 &  +3.207578 &  6.65 & 134.474796 &  +3.176485 &  0.19 & 81.39 & 0.25 & CAMIRA    \\
 16 & 179.703694 & -1.081777 &  6.50 & 179.670749 & -1.066721 &  0.15 & 34.72 & 0.24 & CAMIRA    \\
 17 & 213.761632 & -0.439816 &  6.49 & 213.651488 & -0.406581 &  0.15 & 87.57 & 0.75 & CAMIRA    \\
 18 & 156.740791 &  +0.080817 &  6.49 & 156.794729 &  +0.006912 &  0.34 & 30.61 & 1.10 & redMaPPer \\
 19 & 235.482589 & +42.819316 &  6.45 & 235.487751 & +42.821906 &  0.24 & 48.72 & 0.04 & CAMIRA    \\
 20 & 129.811107 &  +2.449030 &  6.27 & 129.747767 &  +2.483233 &  0.35 & 24.28 & 0.90 & CAMIRA \\
... &  ... &   ... &   ... &  ... &   ... &   ... &   ... &  ... &  ... \\
\enddata

\tablecomments{\autoref{tab:peak_list} is published in its entirety in the machine-readable format\footnote{\url{https://github.com/inonchiu/hsc_shear_selected_clusters}}.
A portion is shown here for guidance regarding its form and content.}
%% No \tablerefs indicated

\end{deluxetable*}

%%%%%%%%%%%%%%%%%%%%%%%%%%%%%%%%%%%%%%%%%%%%%%%%%%
%
% Methodology
%
%%%%%%%%%%%%%%%%%%%%%%%%%%%%%%%%%%%%%%%%%%%%%%%%%%

\section{Modeling Framework and Injection Simulations} \label{sec:method}
One of the biggest challenges in cluster cosmology is to understand and quantify the mass-observable relation. For weak-lensing shear-selected clusters, the most direct observable is the signal-to-noise ratio $\nu$ associated with the peak detected on the mass maps. Other observables, such as cluster redshifts, could be obtained through matching peaks with other cluster catalogs with available redshift information \citep{S16A_S2C,S19A_S2C} or through a dedicated confirmation tool \citep[e.g.,][]{MCMF, MCMF_EFEDS}. In this work, we only consider the cluster number counts as a function of the signal-to-noise ratio $\nu$. 

The number count of weak-lensing peaks as a function of their signal-to-noise ratio $N(\nu)$ is related to the underlying halo mass function through 
\begin{equation}
\label{eq:num_count}
\begin{aligned}
    \frac{\dif N(\nu | \vect{p}) }{\dif \nu}  = \Omega_\mathrm{survey}\iint\,\dif M \dif \redshift&\Bigg[ \frac{\dif n(M | \redshift, \vect{p})}{\dif M} \frac{\dif V(\redshift | \vect{p})}{ \dif\redshift} \\ 
    &\times P( \nu | M, \redshift, \vect{p}) \Bigg],
\end{aligned}
\end{equation}
where $\Omega_\mathrm{survey}$ is the survey area, ${\dif n(M| \redshift, \mathbf{p})}/{\dif M}$ is the halo mass function at a given redshift $\redshift$, $\dif V(\redshift | \vect{p}) / \dif\redshift$ is the differential comoving volume element, and $P( \nu | M, \redshift, \vect{p})$ is the probability of observing $\nu$ given the cluster halo mass $M$ at the redshift $\redshift$. We note that all of these terms are sensitive to cosmology, which is incorporated in the parameter vector $\vect{p}$. A correct prediction of the peak number count therefore heavily depends on the probability distribution $P( \nu | M, \redshift, \vect{p})$, widely referred to as the mass--observable relation. It is precisely the goal of this paper to model the mass--observable relation for these weak-lensing peaks.

A widely adopted analytical model for weak-lensing peak counts was proposed in \citet{Fan:2010} and has been applied to constrain cosmology in \citet{Shan:2012, Shan:2014, Shan:2018} and \citet{Liu:2023}. However, such a model heavily relies on the assumption that various sources of scatter on the weak-lensing observable are Gaussian in nature and cannot probe the full complexity of these systematic uncertainties. \citet{Lin-CA:2015} proposed a semi-analytical model by injecting halo into N-body simulation to quantify the full impact of projection from LSS, intrinsic alignment, and survey masks. In this work, we adopt a more sophisticated semi-analytical model to quantify the mass--observable relation. Instead of painting halos into N-body simulations, we inject the lensing signals of halos into the observed weak-lensing mass maps to determine $P( \nu | M, \redshift, \vect{p})$. This is achieved by injecting the synthetic shear distortions into the ellipticity of the source galaxies observed in the HSC-Y3 shape catalog.

By injecting the signals of synthetic clusters into the observed shape catalogs, the resulting weak-lensing observable naturally accounts for the observed noise and the systematic uncertainties inherent in the real universe. This is because the shape distortion of these source galaxies already encompasses contributions such as shape noise, the projection from uncorrelated LSS, and the intrinsic alignment of source galaxies. Meanwhile, by repeatedly injecting halos at various positions in our survey footprint, we are able to characterize not only the complicated geometry of the survey footprint, including the masking of bright stars, but also the variation in imaging depth. In addition, our approach accounts for any numerical effects associated with mass-map making in the real analysis, such as the flat-sky projection and pixelization.

However, repeating these injection simulations for all possible cosmological parameters $\vect{p}$ to quantify $P( \nu | M, \redshift, \vect{p})$ is computationally challenging. One possible approach is to perform simulations on a grid of cosmological models and leverage an emulator to interpolate among different cosmologies \citep[e.g.][]{Liu_J:2015, Kacprzak:2016, Nishimichi2019}. Such a method is still computationally challenging to scale up with an increasing amount of cosmological and nuisance parameters that are necessary for cluster cosmology analyses.

In this work, we propose a novel approach to dissect the selection function and the mass--observable relation in a physical way and factor out the cosmological dependence in the part that requires the computationally intensive injection simulations. To be more specific, we re-write $P( \nu | M, \redshift, \vect{p})$ into
\begin{equation}
\label{eq:mass-observable}
    \begin{aligned}
        &\,P( \nu | M, \redshift, \vect{p} ) \\
        =& \iint\dif\hat{\aperKappa} \dif\thetas
P( \nu | \hat{\aperKappa}, \thetas)
P(\hat{\aperKappa}, \thetas |\mass, \redshift, \vect{p}) \, .
\end{aligned}
\end{equation}
Here, $\hat{\aperKappa}$ and $\thetas$ are parameters representing the analytic aperture mass and the characteristic angular scale of the cluster, which will be precisely defined later in Sec.~\ref{subsec:reparam}. These parameters capture the shape {and amplitude} of the lensing profile. This way, we divide the mass--observable relation $P( \nu | M, \redshift, \vect{p})$ into two pieces: (1) $P( \nu | \hat{\aperKappa}, \thetas)$ is the re-parametrized scaling relation that connects the shape {and amplitude} of a lensing profile to the final observable $\nu$; and (2) $P(\hat{\aperKappa}, \thetas |\mass, \redshift, \vect{p})$, which converts the cluster's physical properties $\left(\mass, \redshift\right)$ to the shape {and amplitude} of the lensing profile parameterized by $\left(\hat{\aperKappa}, \thetas\right)$.
The first term, $P( \nu | \hat{\aperKappa}, \thetas)$, accounts for the \textit{measurement uncertainties} including shape noise, cosmic shear, shape measurement bias, the variation in survey depth, etc. This needs to be determined through computationally expensive injection simulations. However, given a lensing profile determined by $\left(\hat{\aperKappa}, \thetas\right)$, the process of injecting a synthetic cluster signal, re-detecting it on the resulting mass maps, and quantifying the relation $P( \nu | \hat{\aperKappa}, \thetas)$ is insensitive to cosmological parameters and can be computed in advance. The cosmological dependency goes entirely into the conversion $P(\hat{\aperKappa}, \thetas |\mass, \redshift, \vect{p})$. This term also captures the \textit{intrinsic uncertainties}, such as the intrinsic scatter in the weak-lensing observable due to the diversity of halo profiles (e.g., halo concentration and triaxiality) and the weak-lensing mass bias arising from the inaccurate assumption about the halo profiles. The modeling of $P(\hat{\aperKappa}, \thetas |\mass, \redshift, \vect{p})$ will be examined in-depth in our cosmological analysis \citep{PartII}.

In this paper, we focus on deriving $P( \nu | \hat{\aperKappa}, \thetas)$. In Sec.~\ref{subsec:mock}, we discuss the mock cluster samples used in the injection simulations. Details of the injection simulations are given in Sec.~\ref{subsec:sim}. In Sec.~\ref{subsec:reparam}, we introduce the parameters $\left(\hat{\aperKappa}, \thetas\right)$ and provide an analytical argument on how they determine the lensing profile, thus ensuring the cosmological independence of $P( \nu | \hat{\aperKappa}, \thetas)$.

\subsection{The Mock Cluster Samples} \label{subsec:mock}

We generate mock cluster catalogs analytically and inject their weak-lensing signals into the observed shape catalogs to quantify the mass--observable relation $P( \nu | \hat{\aperKappa}, \thetas)$. With the goal of sampling the parameter space of $\left(\hat{\aperKappa}, \thetas\right)$ as uniform and as complete as possible, we adopt the following sampling strategy: We first create a fine uniform grid on the mass $\Mtwooo$ and redshift $\redshift_\mathrm{cl}$ space. Here, $\Mtwooo$ stands for the mass enclosed within the radius of $r_{200\mathrm{c}}$, where the average density inside is $200$ times the critical density of the universe. For each point on the $\Mtwooo$--$\redshift_\mathrm{cl}$ plane, we create our cluster sample with a uniform sampling on the characteristic angular size $\thetas$ of the cluster. Here, $\thetas$ is linked to the halo concentration parameter in a cosmology-dependent way:
\begin{equation}
\label{eq:rs_thets}
    \thetas \coloneqq \frac{r_s}{D_A(\redshift_\mathrm{cl})};
    \quad\textrm{where }r_s\coloneqq \frac{r_{200\mathrm{c}}}{\Ctwooo},
\end{equation}
and $D_{\mathrm{A}}$ is the angular diameter distance. Note that a fixed set of $(\Mtwooo, \redshift_\mathrm{cl}, \thetas)$ uniquely determines the halo concentration $\Ctwooo$ which in turn fixes the shape of the halo density profile if we assume the spherical \citealt*{NFW} (hereinafter NFW) model. It is worth mentioning that a uniform sampling in the space of $\Mtwooo, \redshift_\mathrm{cl}$, and $\thetas$ does not imply a uniform sampling in $\hat{M}_\kappa$ and $\thetas$. However, the sampling range in $(M_{200 c}, \redshift_\mathrm{cl}, \thetas)$ used to generate mock clusters is large enough to cover the the range of $\left(\hat{M}_\kappa, \thetas\right)$ of interest.

In practice, we sample 2554880 clusters for each connected field in our survey. 
{These clusters are sampled uniformly in the $\log\left[\Mtwooo\right]$--$\redshift_\mathrm{cl}$--$\log\left[\thetas\right]$ parameter space with the boundaries of $5\times10^{12}\,M_\odot h^{-1} \leq \Mtwooo \leq 5\times10^{16}\,M_\odot h^{-1}$, $0.02 \leq \redshift_\mathrm{cl} \leq 2.00$, and $0.01' \leq \thetas \leq 30'$.}

\begin{figure*}[tbh]
    \centering
    \epsscale{1.175}
    \plotone{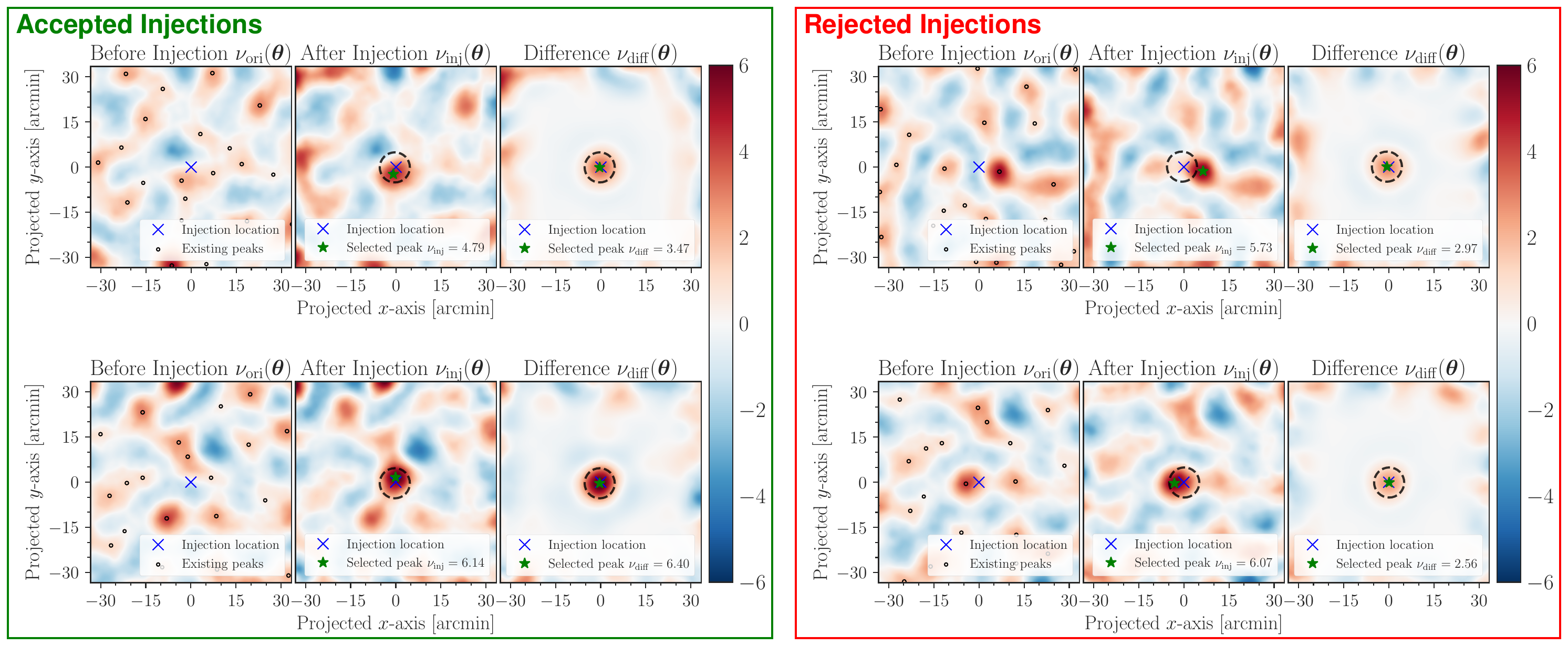}
    \caption{Examples of injecting cluster lensing signal into the HSC-Y1 XMM field shape catalog following the procedure outlined in Sec.~\ref{subsec:sim}. The left and the right panel each shows two cases that are accepted or rejected by step~\ref{it:rejection} in Sec.~\ref{subsec:sim} in order to remove cases where we select peaks that are not mainly associated with the injected halos. The dashed circles in these plots indicate the $5$ arc minute range around $\vect{x}_\mathbf{diff}$ beyond which we reject a detection.\label{fig:before_afer_diff}}
\end{figure*} 

\subsection{Injection Simulations} \label{subsec:sim}
To quantify the mass--observable relation unique to each field in our survey, we inject the weak-lensing signals of mock clusters into the observed shape catalog. For each of the connect fields in our survey and each of the mock clusters, we perform an injection in the following steps:

\begin{enumerate}[label=(\roman*), leftmargin=0pt, itemindent=20pt,
labelwidth=15pt, labelsep=5pt, listparindent=0.7cm,
align=left]
    \item \label{it:i} We randomly assign a position $\vect{\theta}$ in the minimum bounding rectangle of a given field. As a result, mock clusters are injected not only into the region where we observe source galaxies but also into masked areas or areas close to the survey boundary. This is because it is possible to have a cluster where its center is masked, but the outskirts of its lensing profile could still create an off-centered peak in the mass map due to the presence of shape noise and cosmic shear.

    \item \label{it:ii} For each randomly assigned sky position $\vect{\theta}$, we cut out a $4\thetaout\times4\thetaout$ square centered at $\vect{\theta}$. $\thetaout$ denotes the radius of the filter we choose to create the weak-lensing mass maps. Here, we adopt the flat-sky approximation and use the tangent plane projection as described in \citet{S19A_S2C}. Then, for every source galaxy in the cut-out, we assign it a redshift value $z_\mathrm{gal}$ based on a random draw from its photometric redshift distribution $P(z_\mathrm{gal})$. With this, we can calculate the distortion due to the injected cluster on every source galaxy within the patch. 

    \item  \label{it:inject} To calculate the lensing signal, we assume that the mass density profile of the mock cluster follows a spherical NFW profile. 
    \begin{equation}
        \begin{aligned}
            \rho_{\mathrm{NFW}}\left(r\right) &= \displaystyle\frac{ \rho_{\mathrm{s}} }{ 
            \left(r/r_\mathrm{s}\right)\left(1 + r/r_\mathrm{s} \right)^2
            } \\
            \rho_{s} &\coloneqq\frac{200 \rho_{\mathrm{cri}}\left(\clz\right)}{3} \frac{\Ctwooo^{3}}{\ln (1+\Ctwooo)-\Ctwooo /(1+\Ctwooo)} \,.
        \end{aligned}
    \end{equation}    
    The characteristic radius $r_s$ and the concentration parameter $\Ctwooo$ is derived from the given set of $(\Mtwooo, \clz, \thetas)$ and cosmology. It is then straightforward to calculate the shear and the convergence of a source galaxy at redshift $z_\mathrm{gal}$ and the sky position $\vect{\theta}_\mathrm{gal}$:
    \begin{equation}
        \begin{aligned}
            \gamma_\mathrm{NFW}\left(\theta_\mathrm{gal}\right) &= \frac{\Bar{\Sigma}( <\theta_\mathrm{gal}) - \Sigma(\theta_\mathrm{gal})}{\Sigma_\mathrm{cri}(\clz, z_\mathrm{gal}) } \\
            \kappa_\mathrm{NFW}\left(\theta_\mathrm{gal}\right) &= \frac{\Sigma(\theta_\mathrm{gal})}{\Sigma_\mathrm{cri}(\clz, z_\mathrm{gal}) } \, ,
        \end{aligned}
    \end{equation}
    where the projected surface density $\Sigma(\theta)$ is defined as, 
    
    \begin{equation}
    \Sigma\left(\theta\right) \coloneqq \int\mathrm{d}x_3\, \rho_{\mathrm{NFW}}\left(
    \sqrt{
    x_3^2 + \Big(\theta D_{\mathrm{A}}\left(z_{\mathrm{cl}}\right)\Big)^2
    }
    \right), 
    \end{equation}
    and the critical surface density $\Sigma_{\mathrm{cri}}^{-1}(\clz, z_\mathrm{gal})$ is defined as
    \begin{equation}
    \begin{aligned}
        \begin{cases}
        0 &\quad\text{if  } z_\mathrm{gal} \leq \clz,\\
        &\\
        \displaystyle\frac{4\pi G}{c^2}\frac{D_A(\clz)D_A(\clz, z_\mathrm{gal})}{D_A(z_\mathrm{gal})} &\quad\text{otherwise}.
        \end{cases}
    \end{aligned}
    \end{equation}
    The shear on the source galaxy due to the injected halo is expressed in terms of shape measurement $(e_1, e_2)$ as
    \begin{equation}
        e^\mathrm{clu}_{\mathrm{gal}, i} = 2\mathcal{R}_\mathrm{gal}\left[(1+m_\mathrm{gal})\frac{\gamma_{\mathrm{NFW}, i}(\theta_\mathrm{gal})}{1 - \kappa_\mathrm{NFW}(\theta_\mathrm{gal})}\right],
    \end{equation}
    where $\mathcal{R}_\mathrm{gal}$ is the per-galaxy shear responsivity and $m_\mathrm{gal}$ is the per-galaxy calibration bias \citep{S16A_shear}. 
    The resulting shear is then added into the observed ellipticity of the source galaxy. In the weak-lensing limit, the ellipticities of the source galaxies before ($e^\mathrm{before}_{\mathrm{gal}, i}$) and after ($e^\mathrm{after}_{\mathrm{gal}, i}$) the injection is related to each other as
    \begin{equation}
        e^\mathrm{after}_{\mathrm{gal}, i} = e^\mathrm{before}_{\mathrm{gal}, i} + e^\mathrm{clu}_{\mathrm{gal}, i} \, .
    \end{equation}
    We note that while we model the halo lensing signal by assuming a spherical NFW profile here, we will model the deviation from this assumption through a weak-lensing-mass-to-mass scaling relation in \cite{PartII}.
    
    \item \label{it:identify_inj} The shape catalog with the injected weak-lensing signal of the mock cluster is then passed into the pipeline described in Sec.~\ref{subsec:mass_maps} to generate the signal-to-noise ratio map. Following the procedure of constructing the observed peak catalog, we first identify all the positive peaks in the resulting map. We then discard all the peaks that are within a $5$ arcmin radius of any other stronger peaks. The remaining peak that is the closest to the injection center is selected. Therefore, for each of the halo we injected, we obtain two observable, the peak signal-to-noise ratio $\nu_\mathrm{inj}$ and its location $\vect{x}_\mathrm{inj}$.

    \item  \label{it:rejection} Lastly, we need to determine whether the detected peak is mainly associated with the injected halo instead of any existing structures on the weak-lensing map. To address this, we introduce the difference map defined as
    \begin{equation}
        \label{eq:nudiff}
        \nu_\mathrm{diff}(\vect{\theta}) \coloneqq \frac{M^\mathrm{after}_\kappa(\vect{\theta}) - M^\mathrm{before}_\kappa(\vect{\theta})}{\sigma(\vect{\theta})}.
    \end{equation}
    We select the closest peak on the difference map in the same manner as described in step~\ref{it:identify_inj}. The resulting peak on the difference map at the location $\vect{x}_\mathrm{diff}$ with the signal-to-noise ratio $\nu_\mathrm{diff}$ serves as a proxy for the expected signal-to-noise of the injected halo in the absence of cosmic shear and shape noise. With $\left(\vect{x}_\mathrm{inj}, \nu_\mathrm{inj}\right)$ and $\left(\vect{x}_\mathrm{diff}, \nu_\mathrm{diff}\right)$ of the peaks detected in the post-injection and difference maps respectively, we consider the injected halo to be detected if 
    \begin{equation}
    \label{eq:reject_criterion}
        |\vect{x}_\mathrm{inj} - \vect{x}_\mathrm{diff}| \leq 5';\quad |\nu_\mathrm{inj} - \nu_\mathrm{diff}| \leq \Delta_s \, ,
    \end{equation}
    where $\Delta_s$ is a nuisance parameter that is not fixed a priori and needs to be calibrated by the data. 
   
\end{enumerate}

\autoref{fig:before_afer_diff} shows four examples in our injection simulation following the procedure outlined above. For each example, we show the weak-lensing map before and after we inject the mock cluster lensing signal as described in step~\ref{it:inject} together with the difference map introduced in step~\ref{it:rejection}. The right panel demonstrates two cases that are rejected in step~\ref{it:rejection} to avoid detecting peaks that are not mainly associated with the injected halos. The top row shows a rejected case as it fails to meet the $|\vect{x}_\mathrm{inj} - \vect{x}_\mathrm{diff}| \leq 5'$ criterion. The bottom row shows another case that is rejected as it has $|\nu_\mathrm{inj} - \nu_\mathrm{diff}| > \Delta_s$ if we choose $\Delta_s = 3.0$. This does not imply that the detected peak here is completely unassociated with the injected halo. The same case would be considered as a detection if we allow for a larger ranging of scattering and choose $\Delta_s = 4.0$ for instance. Therefore, the allowed range of scattering $\Delta_s$ cannot be determined through injection simulation alone and must be calibrated by leveraging external information. In Sec.~\ref{subsec:validation}, we discuss how $\Delta_s$ affects the number count of weak lensing peaks predicted by our framework and leverage optical counterparts to inform $\Delta_s$. In our cosmological analysis \citep{PartII}, the scatter parameter $\Delta_s$ will be self-calibrated by the number count $N(\nu)$ of the shear-selected clusters.

\begin{table}[tb]
    \centering
    \caption{Different sets of reference cosmological parameters adopted in this study.}\label{tab:cosmo}
    \begin{tabular}{l|l}
    \toprule
        Vanilla & $h = 0.7\,\Omega_m = 0.3,\,\Omega_b = 0.05,\,\Omega_k = 0,$ \\
        & $\sigma_8 = 0.80,\,n_s = 0.95,\,w_0 = -1.0$ \\
        \midrule
        Exotic & $h = 0.7,\,\Omega_m = 0.2,\,\Omega_b = 0.05,\,\Omega_k = 0,$ \\
        & $\sigma_8 = 0.65,\,n_s = 1.90,\,w_0 = -1.5$ \\
        \midrule
        Planck\tablenotemark{a} & $h = 0.673,\,\Omega_m = 0.315,\,\Omega_b = 0.049,\,\Omega_k = 0,$ \\
        & $\sigma_8 = 0.812,\,n_s = 0.9649,\,w_0 = -1.0$ \\
    \bottomrule 
    \end{tabular}
    \tablenotetext{1}{\citet{Planck18}}    
\end{table}

We note that in calculating the lensing signal in step~\ref{it:inject}, we need to specify a certain cosmology. In Sec.~\ref{subsec:reparam}, we show that we can parameterize the mass--observable relation in a way that is independent of the choice cosmological parameters. To verify this, we perform the injection simulations outlined here for three different sets of cosmological parameters. These different sets of parameters are summarized in \autoref{tab:cosmo} and the result is given in Sec.~\ref{subsec:validation}. 

We also note that as we adopt a fully blinded framework for our cosmological analysis by working with three copies of blinded shape catalogs. These injection simulations are carried out for each of the blinded catalog. We utilize unblinded shape catalog in a small subfield (the XMM field) from the previous HSC S16A internal data release \citep[hereinafter the HSC-Y1 data]{shape_bias} to validate our pipeline and our modeling framework. All figures shown in this section and the next are based on injection simulations performed on the HSC-Y1 XMM field. For details of the HSC blinding strategy, we refer the reader to \citet{HSCY3CosmicShearPS, HSCY32ptCF} and Part II of this series \citep{PartII}. 

To summarize, by repeatedly performing the injection simulations, we can obtain a mapping between any cluster properties to their observed signal-to-noise ratio on the weak lensing maps. This allows us to calculate the selection effect as a function of these cluster properties and also derive the mass--observable relation for our cluster sample. In the following, we demonstrate that we can choose two particular parameters to completely describe the cluster lensing profile and parameterize the mass--observable relation in a way that is independent of cosmological parameters. 

\begin{figure*}[tbh]
    \centering
    \epsscale{1.175}
    \plotone{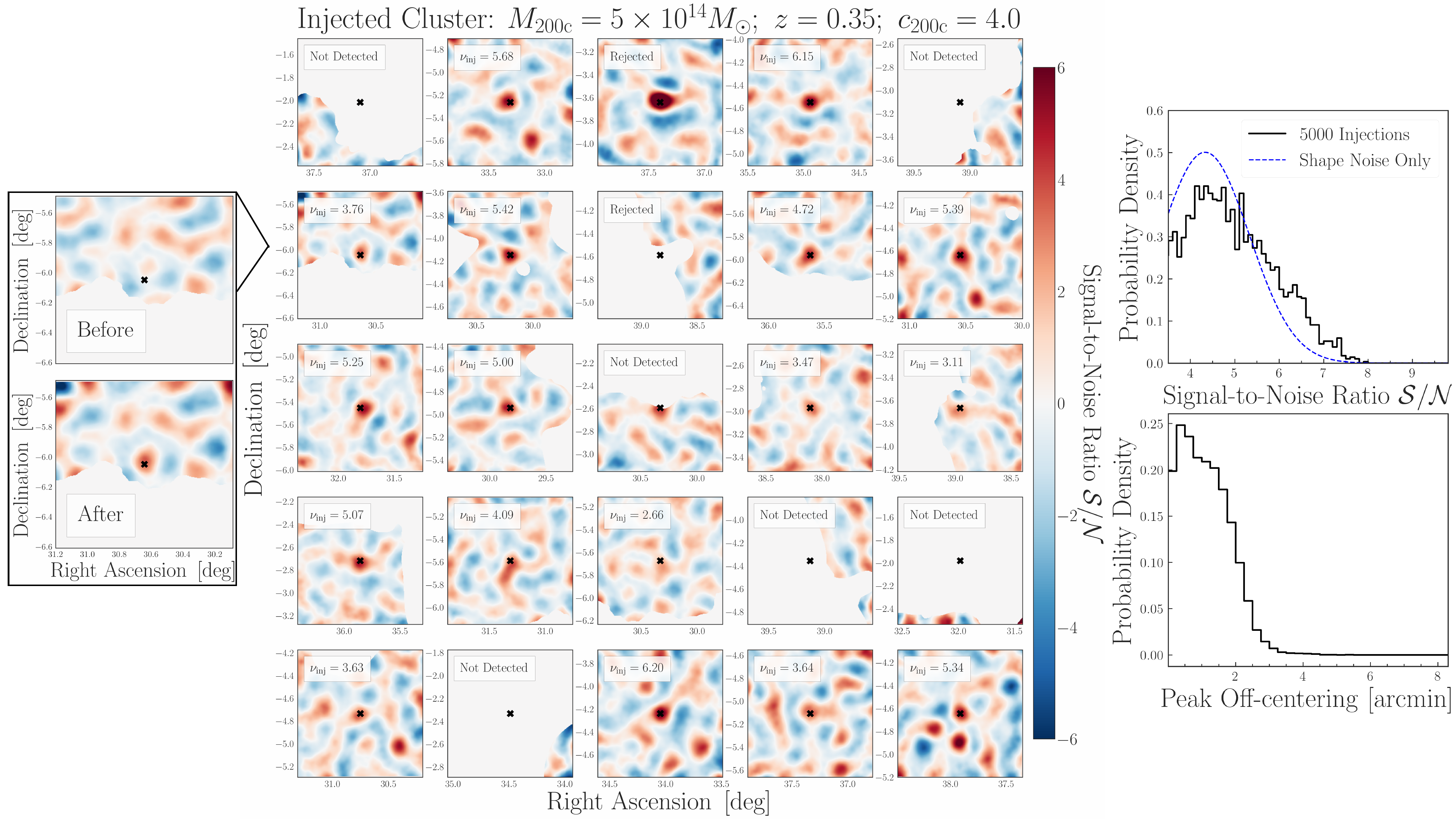}
    \caption{Examples of injecting a single mock halo at various locations on the weak lensing mass map derived from the HSC-Y1 data. The peak singal-to-noise ratio $\nu_\mathrm{inj}$ after each injection is given in the middle panel. Here, \textit{not detected} indicates a peak is not found on the map after we inject the halo, most likely due to masking; and \textit{rejected} indicates a peak is found but does not satisfy the requirement given in step~\ref{it:rejection} in Sec.~\ref{subsec:sim} for us to determine that it is associated with the injected halo. The rightmost two panels show the distributions of peak signal-to-noise ratio $\nu$ and the distance between the observed peak and the halo center from 5000 injections of this same halo. As a comparison, the distribution of peak signal-to-noise ratio predicted from an uniform shape noise is indicated as the dashed blue curve. \label{fig:inject}}
\end{figure*}

\subsection{Re-parametrization of Scaling Relation} \label{subsec:reparam}
The weak-lensing profile of a given cluster involves a calculation that is sensitive to cosmology. However, once we fix the shape of a lensing profile in angular space, the conversion to the shear of the source galaxies, the construction of weak-lensing maps, and the detection of peaks are all independent of cosmological parameters. Thus, the key to parameterize the selection function and the mass--observable relation into a cosmology-independent form is to find parameters that describe the shape of the cluster lensing profile on the weak-lensing map. 

Following Sec.~\ref{subsec:mass_maps}, the lensing signal around a galaxy cluster placed at the origin of a weak-lensing map is given by
\begin{equation}
\begin{aligned}
    \aperKappa(\vect{\theta})&=\iint \dif\vect{\theta^{\prime}} \kappa(\vect{\theta^{\prime}}) U(|\vect{\theta^{\prime}} - \vect{\theta}|) \\
    &=\iint \dif\vect{\theta^{\prime}} \frac{\Sigma(\vect{\theta}^{\prime})}{\Sigma_\textrm{cri}(\vect{\theta}^{\prime})} U(|\vect{\theta^{\prime}} - \vect{\theta}|).
\end{aligned}
\end{equation}
Although we do not know the true redshift of the source galaxies to determine $\Sigma_\textrm{cri}$ and the true halo density profile to calculate $\Sigma$, we can estimate $\aperKappa(\vect{\theta})$ using the ensemble average of $\Sigma_{\mathrm{cri}}^{-1}(\clz, z_\textrm{gal})$ and adopt the lensing signal from an NFW-like halo. The lensing signal from an NFW-like halo on the weak-lensing map is 
\begin{widetext}
\begin{equation}
    \aperKappa(\vect{\theta}) \approx 2\pi\rho_s r_s \left\langle\Sigma_{\mathrm{c}}^{-1}(z_\mathrm{cl})\right\rangle \int_0^{\thetaout} U(\theta - \theta^{\prime})f\left(\frac{D_A(z_{\mathrm{cl}})\theta^{\prime}}{r_s}\right) \theta^{\prime} \dif \theta^{\prime} 
\end{equation}
where $f(x)$,  given by \citet{NFW_lensing}, is
\begin{equation}
f(x) \coloneqq
    \begin{cases}
     \displaystyle\frac{2}{x^2 - 1}\left[1 - \frac{2}{\sqrt{1-x^2}}\operatorname{arctanh}\sqrt{\frac{1-x}{1+x}}\right]\quad x < 1, \\
     \displaystyle\frac{2}{3}\quad x = 1, \\
     \displaystyle\frac{2}{x^2 - 1}\left[1 - \frac{2}{\sqrt{x^2-1}}\operatorname{arctanh}\sqrt{\frac{x-1}{x+1}}\right]\quad x > 1.
    \end{cases}
\end{equation}
Because of the shape of our chosen filter, it was shown that the aperture mass is less sensitive to the change of halo density profiles \citep{Chen2020}. Together with a realistic estimate of $\langle\Sigma_{\mathrm{cri}}^{-1}\rangle$ using source galaxies' photometric redshift distributions, this gives us a faithful representation of a cluster's signature on the weak-lensing maps. If we perform a change of variable to $x\coloneqq D_A \theta^\prime / r_s$ and use $\theta_s = r_s/D_A$, we get
\begin{equation}
\begin{aligned}
    \aperKappa(\vect{\theta}) \approx& 2\pi\rho_s r^3_s D_A^{-2} \langle\Sigma_{\mathrm{cri}}^{-1}\rangle\int_0^{x_\mathrm{out}\coloneqq\thetaout/\thetas}U\left(\theta - x\thetas\right)f(x) x\dif x \\
    \eqqcolon& \hat{\aperKappa}\frac{\displaystyle\int_0^{x_\mathrm{out}}U\left(\theta - x\thetas\right)f(x) x\dif x}{\displaystyle\int_0^{x_\mathrm{out}}U\left(x\thetas\right)f(x) x\dif x}.
\end{aligned} 
\end{equation}
Here, we define $\hat{\aperKappa}$ to be the peak value of the aperture mass profile for an NFW-like halo
\begin{equation}
\begin{aligned}
\label{eq:mkappa}
    \hat{\aperKappa}(M, z, \thetas) &\coloneqq \left\langle\iint \dif\vect{\theta^{\prime}} \kappa_\mathrm{NFW}(\vect{\theta^{\prime}}) U(|\vect{\theta^{\prime}}|) \right\rangle \\
    &= 2\pi\rho_s r^3_s D_A^{-2} \langle\Sigma_{\mathrm{cri}}^{-1}\rangle\int_0^{x_\mathrm{out}\coloneqq\thetaout/\thetas}U\left(x\cdot\thetas\right)f(x) x\dif x,
\end{aligned}
\end{equation}
in which $\rho_s, r_s$ are obtained from $\mass, \redshift, \thetas$ assuming a NFW-like halo, $D_A$ is the angular diameter distance at the halo's redshift, and $\left\langle\Sigma_{\mathrm{cri}}^{-1}\right\rangle$ is obtained by an ensemble average of $\Sigma_{\mathrm{cri}}^{-1}$ using the source galaxy redshift distribution averaged across the entire field where we inject the halo into. 
\end{widetext}

We thus see that the peak value of the estimated aperture mass profile $\hat{\aperKappa}$ and the characteristic angular scale $\thetas$ completely fix the lensing profile for an NFW-like halo. This indicates that the parameter set $\left(\hat{\aperKappa}, \thetas\right)$ is the sufficient statistic for the distribution $P(\nu)$ and motivates us to postulate that $P( \nu | \hat{\aperKappa}, \thetas)$ is insensitive to cosmology. Our claim is validated numerically in Sec.~\ref{subsec:validation}.

%%%%%%%%%%%%%%%%%%%%%%%%%%%%%%%%%%%%%%%%%%%%%%%%%%
%
% Results
%
%%%%%%%%%%%%%%%%%%%%%%%%%%%%%%%%%%%%%%%%%%%%%%%%%%

\section{Results}\label{sec:result}

The injection simulations performed in Sec.~\ref{sec:method} allow us to obtain a mapping between the lensing properties $\left(\hat{\aperKappa}, \thetas\right)$ of a cluster to its observed signal-to-noise ratio $\nu$. In this section, we show results of these injections in Sec.~\ref{subsec:inject_result} and the corresponding selection function and mass--observable relation in Sec.~\ref{subsec:completeness}. In Sec.~\ref{subsec:validation}, we perform validation tests to examine the predicted number counts of weak lensing peaks. As we adopt a fully blinded framework for our cosmological analysis by working with three copies of blinded HSC-Y3 shape catalogs, the results shown in this section are based on the data from a small subfield (the XMM field) in the HSC-Y1 shape catalog. 

\begin{figure*}[t]
    \centering
    \epsscale{1.175}
    \plotone{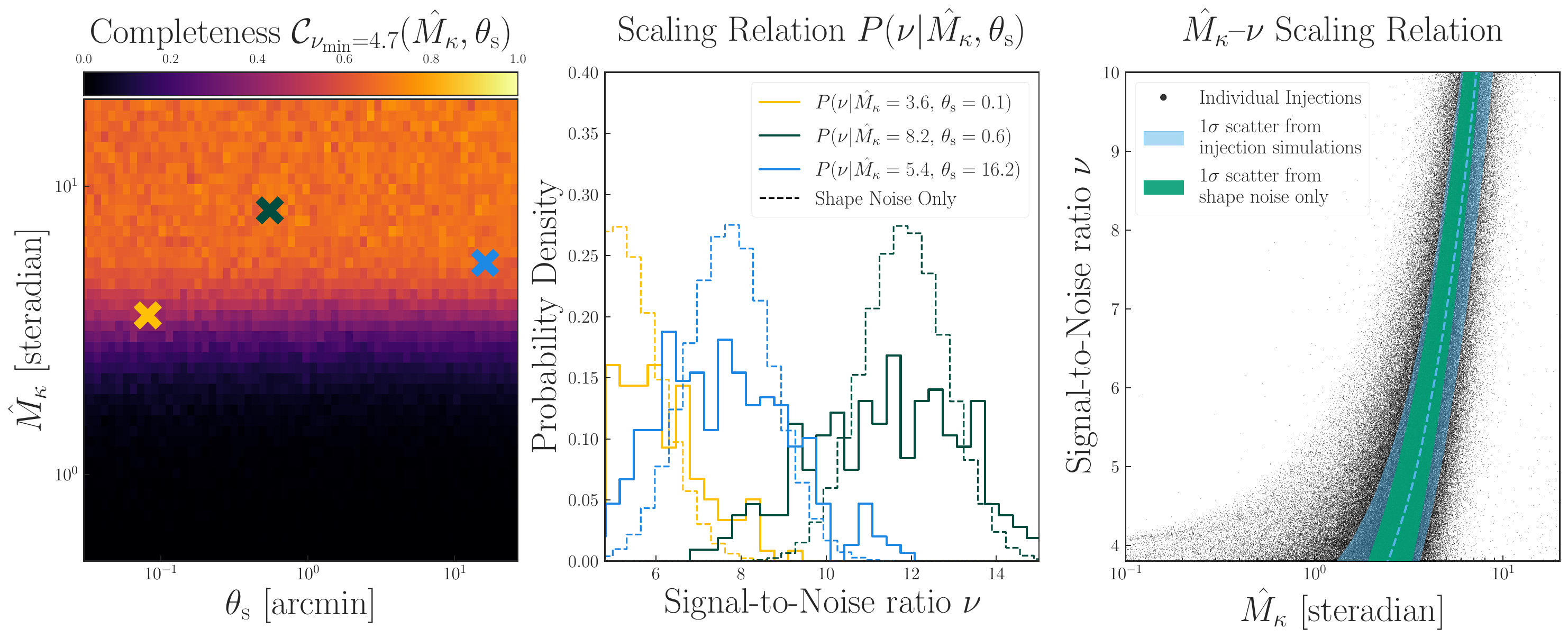}
    \caption{Completeness function and scaling relation derived from injecting clusters into the HSC-Y1 XMM field. From left to right: the completeness function $\mathcal{C}_{\numin}(\hat{\aperKappa}, \thetas)$; the distribution $P(\nu | \hat{\aperKappa}, \thetas)$ at three different points on the $\hat{\aperKappa}$--$\thetas$ plane (indicated as crosses in the leftmost panel); and the marginalized $\hat{\aperKappa}$--$\nu$ scaling relation, with each dot indicating the result from one injection. The shaded regions in the rightmost panel represent the 68.45\% confidence interval of the distribution $P(\nu|\hat{\aperKappa} )$ as a function of $\hat{\aperKappa}$ derived from injection simulations (blue) and from analytical calculations assuming only the shape noise (green). \label{fig:selfunc}}
\end{figure*}

\subsection{Results from Injection Simulations} \label{subsec:inject_result}

\autoref{fig:inject} shows the results of injecting the same cluster into different locations on the weak-lensing mass map. An example of the weak-lensing mass maps before and after the injection is presented in the leftmost panel. Meanwhile, the central panel shows that the same cluster is detected with various signal-to-noise ratios $\nu$ at different locations, revealing significant scatter due to the variations in both the signal from large-scale structures and the shape noise observed across the survey footprint. By performing the injections repeatedly on different locations on the maps, we are able to fully capture the measurement uncertainties in $\nu$ directly from the data without any assumption about the nature of these uncertainties. In some cases, the cluster is not detected due to the survey masking; while in other cases, a peak is found but is rejected due to the criterion in Equation \eqref{eq:reject_criterion}. This means that either the peak has large angular offset to the injected center ($> 5\arcmin$), or the signal-to-noise ratio $\nu$ of the peak is significantly scattered beyond the range we consider ($| \nu_\mathrm{inj} - \nu_{\mathrm{diff}} | > \Delta_s$).

In the rightmost panels in \autoref{fig:inject}, we show the distribution of the signal-to-noise ratio (top panel) and the offset between the detected peak and the injection location (bottom panel). In the top panel, we also show the distribution of $\nu$ assuming only the presence of an uniform and Gaussian shape noise (blue dashed line) as a comparison. This is obtained by a normal distribution centered at $\hat{\aperKappa}(M, z, c) / \sigma$ with $\sigma$ evaluated as \citep{shape_noise, S16A_shear}
\begin{equation}
    \sigma^2 = \frac{\sigma^2_{\epsilon}}{2n_{\mathrm{gal}}}\int_{0}^{\infty}2\pi|Q(\theta)|^2~\theta\mathrm{d}\theta, 
    \label{eq:shape_noise}
\end{equation}
with $\sigma_\epsilon = 0.4$ and $n_\mathrm{gal}$ estimated from the average galaxy number density over the entire HSC-Y1 XMM subfield. While the mean of the two distributions agrees with one another, the signal-to-noise values predicted by the injection simulations exhibit larger scatter. This suggests there are non-negligible contributions to the signal-to-noise ratio $\nu$ from large-scale structures, intrinsic alignments, and the non-uniform survey depth. In the bottom panel, the miscentering distribution shows that the angular offset between the detected peaks and the true centers is mostly at a level of $\lesssim 2\arcmin$. This is expected as the inner radius of the filter we adopted is roughly $2\arcmin$ as shown in the right panel of \autoref{fig:pdz_filter}. In addition, we can obtain the probability of detecting this halo in this subfield at a given signal-to-noise ratio threshold $\numin$ by integrating the distribution $P(\nu)$,
\begin{equation}
    \mathcal{C}_{\numin} \coloneqq \int_{\numin}^\infty P(\nu)\,\dif\nu.
\end{equation}
For this particular halo, we obtain that it has a $\sim33\%$ chance to be detected in a shear-selected sample constructed with $\numin=4.7$. We note that this number depends on the area where we inject halos into. As we inject into an area that is larger than the actual survey footprint, this probability represents the detectability of the cluster multiplied by the ratio of the survey footprint to the area where we inject halos into. 

\subsection{Completeness Function and Scaling Relation}
\label{subsec:completeness}

With the results of the injection simulation for each cluster in our mock catalog, we now obtain the scaling relation $P(\nu | \hat{M}_\kappa, \thetas)$ for the weak-lensing observable. Here, we also define the completeness function for a given selection threshold $\numin$ as
\begin{equation}
    \mathcal{C}_{\numin}(\hat{\aperKappa}, \thetas) \coloneqq \int_{\numin}^{\infty} P(\nu | \hat{\aperKappa}, \thetas) \dif \nu 
\end{equation}
In practice, $P(\nu | \hat{M}_\kappa, \thetas)$ is evaluated by binning the mock clusters in the logarithmic space of $\left(\hat{\aperKappa}, \thetas\right)$ and calculate the probability distribution of the signal-to-noise ratio $\nu$ at each bin.

\autoref{fig:selfunc} shows the selection function and scaling relation 
derived from injection simulations performed over the HSC-Y1 XMM field with the selection $\numin = 4.7$. The leftmost panel shows the selection function $\mathcal{C}(\hat{\aperKappa}, \thetas)$; the middle panel shows the scaling relation $P(\nu | \hat{\aperKappa}, \thetas)$ at three different points on the $\hat{\aperKappa}$--$\thetas$ plane (indicated as crosses in the leftmost panel); and the rightmost panel shows the scatter of $\nu$ as a function of $\hat{\aperKappa}$, with each dot indicating the result from one injection. In the leftmost panel, we observe that the completeness function does not reach unity even for the most massive halos. As discussed before, this is because we inject halos into an area that is larger than the actual survey footprint. The completeness function saturates at the effective coverage of the survey footprint with respect to the injection area.

Similar to \autoref{fig:inject}, the dashed histograms in the middle panel show the distributions of $\nu$ at different $\left(\hat{\aperKappa}, \thetas\right)$ assuming only the effect of an uniform shape noise. We observe that the scatter in $\nu$ from the injection-based results is larger than those of the shape-noise-only distributions, regardless of the quantities $\left(\hat{\aperKappa}, \thetas\right)$. This is more clearly seen in the rightmost panel, where we show the 68.45\% confidence interval of the distribution $P(\nu|\hat{\aperKappa} )$ as a function of $\hat{\aperKappa}$ derived from injection simulations (blue) and from analytical calculations assuming only the shape noise (green). At the high-$\hat{\aperKappa}$ end, halos that have been partly obscured by the survey boundaries or masks could still produce significant peaks to be detected, allowing more down-scatter; at the low-$\hat{\aperKappa}$ end, significantly stronger up-scatter is predicted by the injection simulations due to the presence of large-scale structures. 

\subsection{Validations} \label{subsec:validation}
In this work, we derive the scaling relation $P\left(\nu | \hat{\aperKappa}, \thetas\right)$ directly from data to account for measurement uncertainties on the weak-lensing observable $\nu$. In what follows, we perform several validation tests on our modeling framework by comparing cluster number counts from both data and simulation.

Following Equation \eqref{eq:num_count} and \eqref{eq:mass-observable}, the number counts of shear-selected clusters as a function of the signal-to-noise ratio $\nu$ are calculated as
\begin{equation}
\begin{aligned}
\label{eq:number_count_model}
    &\frac{\dif N(\nu | \vect{p} ) }{\dif \nu}  = \Omega_\mathrm{survey}\iint \dif M \dif \redshift \Bigg[ \frac{\dif n(M | \redshift, \vect{p})}{\dif M} \frac{\dif V(\redshift | \vect{p})}{\dif\redshift} \\ 
    &\times\iint\dif\hat{\aperKappa} \dif\thetas
P( \nu | \hat{\aperKappa}, \thetas)
P(\hat{\aperKappa}, \thetas |\mass, \redshift, \vect{p}) \Bigg]\,,
\end{aligned}
\end{equation}
where $\dif n / \dif \mass$ is modeled by the halo mass function given in \citet{Bocquet16} and $P( \nu | \hat{\aperKappa}, \thetas)$ is obtained by inject simulations performed over a $46.37\,\deg^2$ area that includes the HSC-Y1 XMM footprint. While the actual survey footprint is smaller, since $P( \nu | \hat{\aperKappa}, \thetas)$ already contains information about the effective coverage of the survey footprint in the injection area, $\Omega_\mathrm{survey}$ is chosen accordingly to be $46.37\,\deg^2$. The cosmological parameters $\vect{p}$ are fixed to the values measured by \citet{Planck18}, which is given in \autoref{tab:cosmo}.

The last component to be specified is $P(\hat{\aperKappa}, \thetas |\mass, \redshift, \vect{p})$, the cosmology-dependent conversion from halos' physical properties $(M, z)$ to their lensing properties $\left(\aperKappa, \thetas\right)$. This term will be comprehensively modeled in our cosmological analysis \citep{PartII} to self-consistently take into account the intrinsic uncertainties associated with the halo. For the purpose of validation, we adopt a simplified modeling framework and write
\begin{multline}
    P(\hat{\aperKappa}, \thetas | \mass, \redshift) 
    = \\
    \int \dif\Mwl\, P(\Mwl | \mass, \redshift) P(\hat{\aperKappa}, \thetas | \Mwl, \redshift),
\end{multline}
where we introduce the distribution $P(\Mwl | \mass, \redshift)$ to capture the bias and uncertainties arose from the inaccurate assumption about the halo density profile due to the presence of, e.g., the halo triaxiality \citep[e.g.,][]{Becker2011,Hamana2012}. Here, we assume a redshift-independent model for $P(\Mwl | \mass, \redshift)$ where we consider $\ln\Mwl\sim\mathcal{N}(\ln\mass, 0.18^2)$ \citep{Chen2020}. Meanwhile, we have 
\begin{equation}
    P(\hat{\aperKappa}, \thetas | \Mwl, \redshift) = P(\hat{\aperKappa} | \thetas , \Mwl, \redshift) P(\thetas | \Mwl, \redshift),
\end{equation}
where $P(\hat{\aperKappa} | \thetas , \Mwl, \redshift)$ is a Dirac delta function centered at $\hat{\aperKappa}(\Mwl, \redshift, \thetas)$ given by Equation \eqref{eq:mkappa}, and $P(\thetas | \Mwl, \redshift)$ links the angular size of a halo to its weak lensing mass \Mwl at a given redshift, which we model through the mass--concentration relation. Here, we assume that the mass--concentration relation follows a log-normal distribution around the mean value predicted by \citet{Diemer19} with a scatter $\sigma_{\ln c} = 0.30$. 

\begin{figure}[t]
    \centering
    \epsscale{1.175}
    \plotone{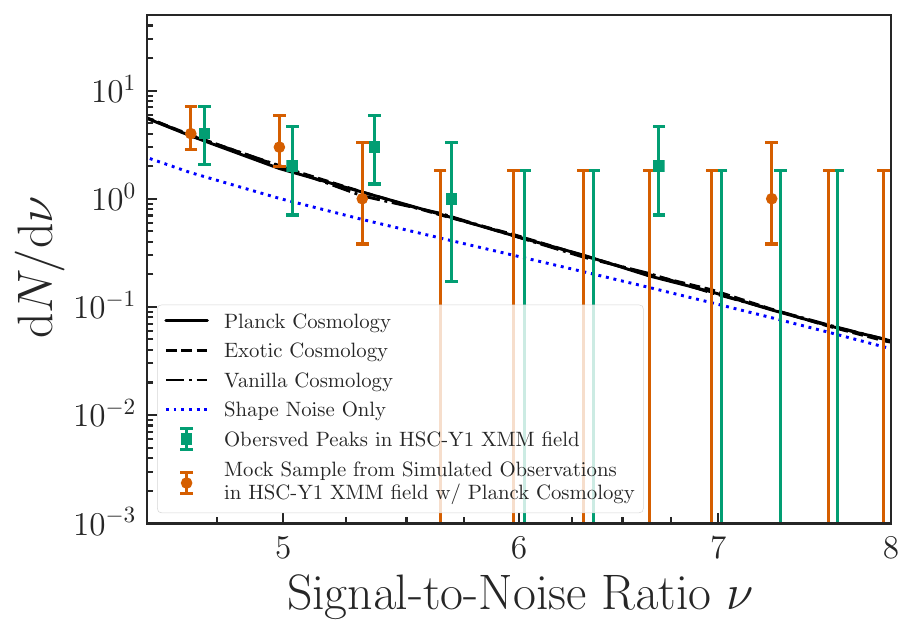}
    \caption{Validation test on the invariance of the scaling relation $P(\nu | \hat{\aperKappa}, \thetas)$ derived under different reference cosmologies. The three black lines (solid, dashed, dash-dotted) show the number counts predicted for the HSC-Y1 XMM field using $P(\nu | \hat{\aperKappa}, \thetas)$ derived under three different sets of cosmological parameters while the dotted blue line is the prediction considering only an uniform shape noise. These results are compared to the number counts in a mock cluster samples (orange circles) and the number counts of real observed peaks (green squares) in this field. The mock sample is constructed through drawing a Poisson realization of the halo mass function and directly inject them into the HSC-Y1 shape catalog to determine which of them can be observed (see Sec.~\ref{subsec:consistency}). For visualization purpose, the green data points are shifted to the right by 0.05 to avoid overlap. \label{fig:validation}}
\end{figure}

\subsubsection{Cosmological Independence}\label{subsec:cosmos_indep}
One of the key results in this work is that the scaling relation $P( \nu | \hat{\aperKappa}, \thetas)$, which accounts for the measurement uncertainties, is independent of the assumed cosmology in which it is derived. Since a rigorous model for the measurement uncertainties requires computationally intensive injection simulations, this allows us to compute $P( \nu | \hat{\aperKappa}, \thetas)$ in advance and apply them to the cosmological analysis in \citet{PartII}.

To validate this, we repeat the injection simulations in Sec.~\ref{subsec:sim} under three different sets of cosmological parameters $\vect{p}_i$ and use the resulting $P(\nu | \hat{\aperKappa}, \thetas, \vect{p}_i)$ to predict number counts
\begin{equation}
\begin{aligned}
    &\frac{\dif N_i(\nu)}{\dif \nu}  = \Omega_\mathrm{survey}\iint \dif M \dif \redshift \Bigg[ \frac{\dif n(M | \redshift, \vect{p})}{\dif M} \frac{\dif V(\redshift | \vect{p})}{\dif\redshift} \\ 
    &\times\iint\dif\hat{\aperKappa} \dif\thetas
P( \nu | \hat{\aperKappa}, \thetas, \vect{p}_i)
P(\hat{\aperKappa}, \thetas |\mass, \redshift, \vect{p}) \Bigg]\,.
\end{aligned}
\end{equation}
Here, $\vect{p}$ is fixed to the Planck cosmology to compute terms that are known to depend on cosmology, while $\vect{p}_i$ represents the three sets of cosmological parameters given in \autoref{tab:cosmo} that is used to compute $P(\nu | \hat{\aperKappa}, \thetas)$. The number counts predicted by the three different $P(\nu | \hat{\aperKappa}, \thetas, \vect{p}_i)$ are shown in \autoref{fig:validation}. As can be seen, the resulting cluster number counts $\dif N_i/\dif\nu$ are in excellent agreement with one another, strongly demonstrating that the scaling relation $P(\nu | \hat{\aperKappa}, \thetas)$ is insensitive to the underlying cosmology. This offers great computational advantages as it allows us to compute $P (\nu | \hat{\aperKappa}, \thetas)$ in advance and we will only need to compute the other cosmology-dependent terms in Equation \eqref{eq:number_count_model} in the cosmological analysis. 

\subsubsection{Consistency in the Modeling Framework}\label{subsec:consistency}
To ensure the self-consistency and the correctness of our modeling framework, we further compare the predicted number counts from Equation \eqref{eq:number_count_model} with the number counts obtained from a mock observation and from the real observed catalog. 

First, we test whether our modeling framework can recover the number counts of a mock sample generated using a set of known cosmological and nuisance parameters. The mock catalog is created as a Poisson realization of the halo mass function \citep{Bocquet16} up to redshift 2 over an area of $\Omega_\mathrm{survey} = 46.37,\deg^2$. Each mock cluster is assigned a weak-lensing mass, $\Mwl$, with respect to its true mass following a log-normal distribution, $\ln\Mwl\sim\mathcal{N}(\ln\mass, 0.18^2)$. Similarly, the concentration of each mock cluster is randomly drawn from a log-normal distribution with a mean value predicted by \citet{Diemer19} and a scatter of $\sigma_{\ln c} = 0.30$. Following step~\ref{it:ii}-\ref{it:identify_inj} in Sec.~\ref{subsec:sim}, these mock clusters are then injected into the XMM subfield of the HSC-Y1 shape catalog, with the detected clusters forming a mock sample of shear-selected peaks. 

The number counts of the mock sample is shown as orange circles in \autoref{fig:validation}. The excenllent agreement between the number counts from a mock observations and those predicted by Equation \eqref{eq:number_count_model} indicates that the modeling framework we develop is self-consistent. Moreover, we compare these results directly with the number counts of shear-selected clusters observed in this field (green squares). The good agreement (at a level of $\lesssim1\sigma$) seen in \autoref{fig:validation} suggests that the modeling framework provides a good description of the data. 

\begin{figure}[tbh]
    \centering
    \epsscale{1.175}
    \plotone{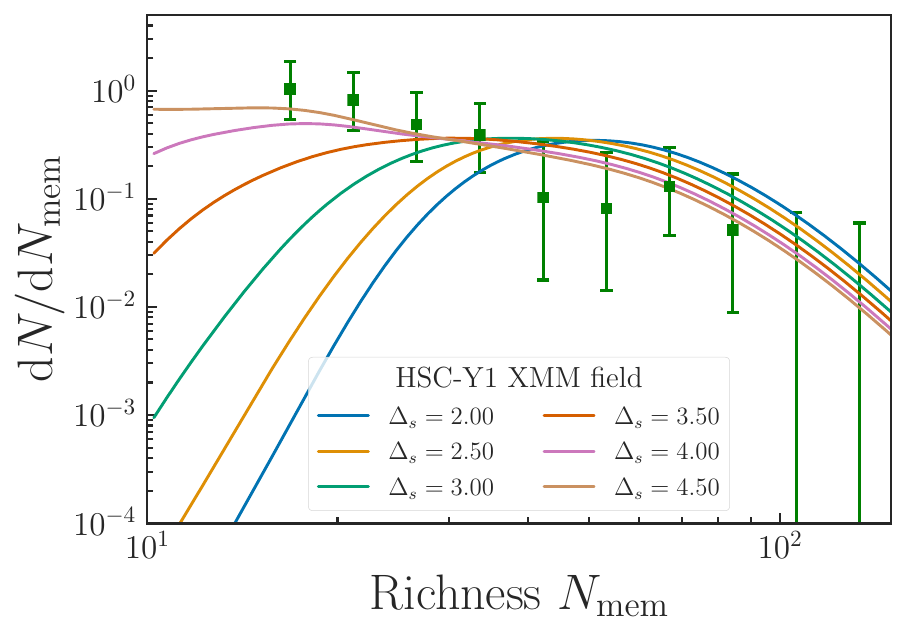}
    \caption{Validation test on the shape of the predicted number counts of weak lensing peaks' optical counterparts as a function of their optical richness given the scaling relation $P(\nu | \hat{\aperKappa}, \thetas)$ derived with different allowed range of scattering $\Delta_s$ (see step~\ref{it:rejection} in Sec.~\ref{subsec:sim}). These predictions are compared with the number counts of the closest CAMIRA counterpart for weak lensing peaks found in the HSC-Y1 XMM field with a signal-to-noise ratio larger than 4.7 (green squares). A statistically more constraining result using the entire HSC-Y3 field obtained after unblinding the cosmological analysis in \citet{PartII} is shown in \autoref{fig:richness_compare-unblind} in Appendix~\ref{appendix:counterpart}. \label{fig:richness_compare}}
\end{figure}

\subsubsection{Comparison with Cross-matched Optical Clusters}\label{subsec:richness_distribution}

Lastly, we compare the richness distributions between the observed shear-selected clusters' optical counterparts and the theoretical prediction using a calibrated richness--mass relation. Based on the optical richness $N_{\mathrm{mem}}$, this comparison provides not only an end-to-end test on the modeling framework but also a constraint on the scattering parameter $\Delta_\mathrm{s}$ used in the injection simulations (see Sec.~\ref{subsec:sim}).

Similar to Equation \eqref{eq:number_count_model}, we can write the number counts of optical counterparts as a function of their richness as
\begin{multline}
    \label{eq:richness_count}
        \frac{ \dif N\left(N_{\mathrm{mem}}\right) }{\dif N_{\mathrm{mem}}}  = 
        \iiint   \Theta \left(\nu - \nu_{\mathrm{min}}\right)
        P\left(N_{\mathrm{mem}}, \nu| \mass, \redshift \right)
        \times \\
        \frac{ \dif N\left(\mass, \redshift | \vect{p}\right) }{ \dif \mass \dif \redshift } 
        \dif\nu\dif\mass\dif\redshift \, ,
\end{multline}
where $\Theta \left(\nu - \nu_{\mathrm{min}}\right)$ is the Heaviside step function that denotes the selection threshold, and the probability $P\left(N_{\mathrm{mem}}, \nu| \mass, \redshift \right)$ describes the joint distribution of $N_{\mathrm{mem}}$ and $\nu$ at a given cluster mass \mass\ and redshift \redshift.
To first-order approximation, we assume that the optical richness $N_{\mathrm{mem}}$ to be independent of the weak-lensing signal-to-noise ratio $\nu$, 
\begin{equation}
    P\left(N_{\mathrm{mem}}, \nu| \mass, \redshift \right) = P\left(N_{\mathrm{mem}} | \mass, \redshift \right) P\left(\nu| \mass, \redshift \right) \, .
\end{equation}
We make use of the richness--mass scaling relation from \citet{Murata2019} to evaluate $P\left(N_{\mathrm{mem}} | \mass, \redshift \right)$. Together with the scaling relation $P(\nu | \hat{\aperKappa}, \thetas)$ derived from the injection simulations, we can compute the predicted richness distribution of the shear-selected clusters. When evaluating Equation \eqref{eq:richness_count}, we note that we restrict the redshift integral to be $\redshift \geq 0.1$, as we compare the richness distribution with that of the optical counterparts from CAMIRA which only covers $z \geq 0.1$. 

On the other hand, we cross-match the weak-lensing peaks observed in the HSC-Y1 XMM field with optical clusters that have $N_\mathrm{mem} \geq 15$ in the CAMIRA catalog. An optical counterpart can be identified within a 5\arcmin\ radius around all peaks with $\nu \geq 4.7$. This gives the shear-selected clusters an observed richness distribution that we can compare with. Note that in this comparison the counterparts are obtained by a simple matching in the sky coordinate and do not utilize the information of richness and redshift of the optical clusters as done in Sec.~\ref{subsec:cl_sample} and Appendix~\ref{appendix:counterpart}. By doing so, the optical counterparts are determined without any prior knowledge of their richness distribution, enabling an independent assessment on the resulting weak-lensing scaling relation and selection function by comparing the richness distribution of the counterparts.

\autoref{fig:richness_compare} shows the comparison between the observed richness distribution and the theoretical predictions using the scaling relation $P(\nu | \hat{\aperKappa}, \thetas)$ derived with different allowed range of scattering $\Delta_s$ (see step~\ref{it:rejection} in Sec.~\ref{subsec:sim}). This plot is generated with the cosmological parameters fixed to those measured by \textit{Planck}. The most distinct feature of changing the parameter $\Delta_s$ is revealed by the overall shape of $\dif N/\dif N_\mathrm{mem}$: The smaller (larger) value of $\Delta_s$ results in a decreasing (increasing) number of clusters at the low-richness end. This is expected, given that the more low-mass systems would be up-scattered into the shear-selected sample with an increasing $\Delta_s$. We see that the observed richness distribution in the HSC-Y1 XMM field (green squares) {favors a higher value in $\Delta_s$}. A statistically more constraining result using the entire HSC-Y3 field obtained after unblinding the cosmological analysis in \citet{PartII} is shown in \autoref{fig:richness_compare-unblind} in Appendix~\ref{appendix:counterpart}.

We stress that the normalization and shape of $\dif N/\dif N_\mathrm{mem}$ are also affected by the cosmological parameters (e.g., $\Omega_{\mathrm{m}}$ and $\sigma_8$) and the scatter in both the richness--mass scaling relation and the weak-lensing mass bias. Therefore, we must simultaneously constrain the parameter $\Delta_s$ together with all the cosmological and nuisance parameters in a self-calibrating way. Moreover, because the optical counterparts shown in \autoref{fig:richness_compare} are determined by only the matching in the sky coordinate, we expect some falsely matched counterparts at the low-richness end compared to that obtained from the more advanced matching including both the cluster redshift and richness (See Appendix~\ref{appendix:counterpart}). It is feasible but challenging to include the richness distribution of the correctly matched counterparts into the data vector and self-consistently constrain the parameter $\Delta_s$. We therefore do not attempt to determine $\Delta_s$ here and leave it to future work. In \citet{PartII}, $\Delta_s$ will be constrained together with other parameters based on the number counts $N\left(\nu\right)$ only.

\subsection{Discussions} \label{subsec:discussion}

In previous studies, weak-lensing peaks are usually modeled either with an analytical halo model or with numerical simulations. An analytical framework models weak-lensing peaks as halos on top of a Gaussian random field, and the selection function is assumed to be a complementary error function \citep[e.g.][]{Kruse1999, Kruse2000, Bartelmann2002, Hamana2004, Fan:2010, S16A_S2C}. These methods are similar to the ``shape noise only'' case in our analysis. While in some studies \citep{Fan:2010, S16A_S2C}, large-scale structures were also taken into account in addition to shape noise, it was still assumed that fluctuations due to large-scale structures are Gaussian. Therefore, these results are similar to our ``shape noise only'' case with slightly higher noise level. However, our injection simulation demonstrates that the fluctuation in the signal-to-noise ratio of a peak is non-Gaussian (see \autoref{fig:inject} and \ref{fig:selfunc}). We attribute this to the non-linear growth of large-scale structures and the variation in shape noise due to the changing survey depth. As a result, we see in \autoref{fig:validation} that the predicted number counts from the shape-noise-only case deviate from the injection simulation at lower signal-to-noise ratios as the impact of non-Gaussian fluctuations becomes stronger.

On the other hand, these non-Gaussian fluctuations can also be fully captured through numerical simulations \citep[e.g.][]{Hamana2004, Hennawi:2005, Wang2009, Kratochvil2010, Yang2011}. However, N-body and ray-tracing simulations are expensive to conduct and are usually only carried out for a limited number of parameter sets. To partly circumvent this challenge, \citet{Lin-CA:2015} developed a ``fast simulation'' by randomly injecting analytical halo profiles into N-body simulations. Even though these randomly injected halos do not correlate with the underlying N-body simulations, \citet{Lin-CA:2015} showed that this semi-analytical approach can faithfully 
reproduce the peak abundance. The formalism developed in this work is similar in spirit but different in two major ways: 1) In our work, instead of injecting analytical halo profiles into N-body simulations, we inject these halos into the observed shear catalog. This allows us to estimate the line-of-sight scattering according to the observed underlying structures of our universe instead of relying on numerical prescriptions. Moreover, we calculate injected weak-lensing signal on individual galaxies which enables us to directly account for fluctuations due to shape noise. 2) Instead of drawing halos from a halo mass function to directly infer the number counts of weak lensing peaks, we uniformly sample halos across a large parameter space to focus exclusively on deriving the scaling relation and the selection function. The number counts are obtained by convolving our results with a halo mass function (see Equation \eqref{eq:number_count_model}).
By combining with the parametrization scheme introduced in Sec.~\ref{subsec:reparam}, our injection simulation only needs to be performed once to capture the measurement uncertainties. Other systematic effects such as photometric redshift uncertainties and deviations from the spherical NFW profile can be easily incorporated during the modeling stage (see Sec.~\ref{subsec:validation} and \citealt{PartII}).

Even with full N-body and ray-tracing simulations, additional systematic effects related to cluster member galaxies and baryonic physics still need to be considered. The main systematic effects on peak statistics are dilution from cluster member galaxies, intrinsic alignment of satellite galaxies in cluster halos, and baryonic effects on the density profile. These systematic effects bias the peak signal-to-noise ratio and therefore affect cosmological constraints. Previous studies often made explicit corrections to these systematic effects that are around $10\%$ \citep{Applegate2014, Melchior2015, Kacprzak:2016}. In our work, we adopt a conservative strategy instead to generate a contamination-free weak-lensing observable. \citet{S19A_S2C} carefully examined different choices of filter shape and source galaxy redshift cut. It was shown that the combination of the filter shape adopted in this work and a source redshift cut of $0.7$ can effectively remove the contributions from cluster member galaxies. With this combination, the source galaxy number density around galaxy clusters in the region where the filter is non-zero matches that of the random field \citep[see Fig.\,9 in][]{S19A_S2C}. This ensures that the weak-lensing maps we generate do not contain dilution from cluster member galaxies and intrinsic alignments of galaxies around the peak. Alignments of source galaxies with other structures along the line-of-sight are taken into account through the injection simulation as we inject signals into the observed shear catalog which already contains the effect of intrinsic alignments. Our adopted filter also removes contributions from cluster centers so that we are less sensitive to uncertainties in the modeling of cluster cores. The remaining uncertainties in halo density profiles are captured by a weak-lensing mass bias relation \citep{Grandis:2021, Chiu:2022} that is self-consistently calibrated alongside our cosmological constraints \citep{PartII}.

%%%%%%%%%%%%%%%%%%%%%%%%%%%%%%%%%%%%%%%%%%%%%%%%%%
%
% Conclusions
%
%%%%%%%%%%%%%%%%%%%%%%%%%%%%%%%%%%%%%%%%%%%%%%%%%%

\section{Conclusions} \label{sec:conclusion}
We have presented the construction of a shear-selected cluster sample through identifying peaks on weak-lensing aperture mass maps derived from the HSC-Y3 shear catalog. Thanks to the exquisite depth of the HSC survey, we have selected source galaxies far in the background (with significant probability of being above redshift $0.7$) when deriving our weak-lensing maps to reduce the dilution of lensing signal from cluster member galaxies. We have also chosen a filter that minimizes contributions from the center of the cluster and from correlated large-scale structures. These choices have made our cluster weak-lensing observable less sensitive to various bias associated with weak-lensing peaks known in the literature. Over $\sim 510\,\mathrm{deg}^2$, we have identified $129$ weak-lensing peaks with signal-to-noise ratio $\nu \geq 4.7$. We have cross-matched these peaks to optical cluster catalogs and confirmed that $\geq 99\%$ of peaks are associated with galaxy clusters. 

To utilize this cluster sample for cosmological studies, we have performed semi-analytical simulations to comprehensively model the mass--observable relation and the selection effect on this cluster sample. We have injected weak-lensing signal of NFW-like halos into the observed shape catalog to capture measurement uncertainties from our actual survey. This method allows us to simulate real-world weak lensing systematic effects such as the line-of-sight large-scale structure, intrinsic alignment of source galaxies, and the miss-centering of weak-lensing peaks. It also enables us to accurately characterize the impact of the complex survey geometry and non-uniform survey depth. For each of the fields in the HSC survey, we have injected halos across a wide range of physical properties to derive the mass--observable relation as a function these cluster properties. We have repeated this process for three doubly-blinded shape catalogs.

As these semi-analytical simulations are still computationally intensive to perform, we have developed a novel parametrization scheme to arrange the weak-lensing scaling relation in terms of cluster lensing properties. If one has chosen some parameters that can capture the shape of the cluster's weak-lensing profile in the angular space, the mapping between weak-lensing profile to the weak-lensing observable will only be affected by observational uncertainties that are insensitive to cosmological parameters. This allows us to compute the weak-lensing scaling relation due to complex observational uncertainties in advance and greatly enhances our ability to characterize the shear-selected cluster sample in a rigorous but computationally feasible way. In this work, we have chosen to parameterize the selection function and the mass--observable relation with the analytic cluster aperture mass $\hat{M_\kappa}$ and the characteristic angular scale of the cluster $\thetas$. An analytical argument have been given in Sec.~\ref{subsec:reparam} to illustrate why these two parameters are sufficient to describe the cluster lensing profile. The cosmology-dependent conversion between cluster physical properties such as mass and redshift to these lensing properties will be presented in \citet{PartII}.

We have validated the weak-lensing scaling relation derived from our injection scheme is indeed insensitive to changes of cosmological parameters. We have derived three sets of scaling relations under three reference cosmologies (\autoref{tab:cosmo}). The resulting weak-lensing peak number counts derived from these scaling relations have been shown to agree with each other excellently (\autoref{fig:validation}). We have also compared these model predictions with the number counts of a mock sample obtained by directly injecting a realization of halo catalog in the weak-lensing mass maps. These tests have demonstrated the self-consistency of our modeling framework. The correctness of our model has been further verified by comparing our predictions with observed number counts of shear-selected clusters as a function of both weak-lensing signal-to-noise ratio and richness of the optical counterparts. These validation tests have been carried out using a small subfield of the HSC-Y1 shear catalog so that we still respect the blinded analysis. 

Our results enable the cosmological study which will be presented in \citet{PartII}.
In \citet{PartII}, we will discuss our cosmological pipeline that models uncertainties arisen from photometric redshift, and the deviation of our analytical description of the halo profiles from true halo properties such as halo triaxiality. The cosmological constraints derived from these shear-selected clusters will shed light on the $S_8$ tension by serving as a consistent check with other cosmological constraints obtained from weak lensing. Looking forward, the methodology developed in this paper can be applied to weak-lensing data from Stage-IV surveys \citep{LSST, Euclid, Roman} to serve as a strong complement to cosmic shear as we further explore the non-Gaussian information from these weak lensing data. 

%%%%%%%%%%%%%%%%%%%%%%%%%%%%%%%%%%%%%%%%%%%%%%%%%%
%
% Acknowledgements
%
%%%%%%%%%%%%%%%%%%%%%%%%%%%%%%%%%%%%%%%%%%%%%%%%%%

\section*{Acknowledgements}
{We thank an anonymous referee for valuable comments and suggestions.} K.-F.C. acknowledges support from the Taiwan Think Global Education Trust Scholarship and the Taiwan Ministry of Education's Government Scholarship to Study Abroad. K.-F.C. thanks the computing resources provided by the Academia Sinica Institute for Astronomy and Astrophysics in Taiwan which are maintained by Dr.~Bau-Ching Hsieh and others. K.-F.C. thanks the hospitality of Adrian Liu at McGill University where this work is carried out in part. I-Non Chiu acknowledges the support from the National Science and Technology Council in Taiwan (Grant NSTC 111-2112-M-006-037-MY3) and the computing resources provided by the National Center for High-Performance Computing (NCHC) in Taiwan. YTL acknowledges supports by the grant NSTC 112-2112-M-001-061. This work was supported by JSPS KAKENHI Grant Numbers JP20H05856, JP22H01260, JP22K21349.

The Hyper Suprime-Cam Subaru Strategic Program (HSC-SSP) is led by the astronomical communities of Japan and Taiwan, and Princeton University.  The instrumentation and software were developed by the National Astronomical Observatory of Japan (NAOJ), the Kavli Institute for the Physics and Mathematics of the Universe (Kavli IPMU), the University of Tokyo, the High Energy Accelerator Research Organization (KEK), the Academia Sinica Institute for Astronomy and Astrophysics in Taiwan (ASIAA), and Princeton University.  The survey was made possible by funding contributed by the Ministry of Education, Culture, Sports, Science and Technology (MEXT), the Japan Society for the Promotion of Science (JSPS),  (Japan Science and Technology Agency (JST),  the Toray Science Foundation, NAOJ, Kavli IPMU, KEK, ASIAA,  and Princeton University.

The Pan-STARRS1 Surveys (PS1) and the PS1 public science archive have been made possible through contributions by the Institute for Astronomy, the University of Hawaii, the Pan- STARRS Project Office, the Max Planck Society and its participating institutes, the Max Planck Institute for Astronomy, Heidelberg, and the Max Planck Institute for Extraterrestrial Physics, Garching, The Johns Hopkins University, Durham University, the University of Edinburgh, the Queen’s University Belfast, the Harvard-Smithsonian Center for Astrophysics, the Las Cumbres Observatory Global Telescope Network Incorporated, the National Central University of Taiwan, the Space Telescope Science Institute, the National Aeronautics and Space Administration under grant No. NNX08AR22G issued through the Planetary Science Division of the NASA Science Mission Directorate, the National Science Foundation grant No. AST-1238877, the University of Maryland, Eotvos Lorand University (ELTE), the Los Alamos National Laboratory, and the Gordon and Betty Moore Foundation.

This paper is based on data collected at the Subaru Telescope and retrieved from the HSC data archive system, which is operated by the Subaru Telescope and Astronomy Data Center (ADC) at NAOJ. Data analysis was in part carried out with the cooperation of Center for Computational Astrophysics (CfCA), NAOJ. We are honored and grateful for the opportunity of observing the Universe from Maunakea, which has the cultural, historical and natural significance in Hawaii.

This paper makes use of software developed for Vera C. Rubin Observatory. We thank the Rubin Observatory for making their code available as free software at \url{http://pipelines.lsst.io/}. Numerical calculations are performed through the \texttt{Python} packages \texttt{numpy} \citep{numpy} and \texttt{scipy} \citep{scipy}. Many of the cosmological and astrophysical calculations in this work rely on the routines wrapped up in \texttt{colossus} \citep{Colossus} and \texttt{astropy} \citep{astropy}. Plots are made available thanks to \texttt{matplotlib} \citep{matplotlib} and \texttt{seaborn} \citep{seaborn}.

%%%%%%%%%%%%%%%%%%%%%%%%%%%%%%%%%%%%%%%%%%%%%%%%%%
%
% References
%
%%%%%%%%%%%%%%%%%%%%%%%%%%%%%%%%%%%%%%%%%%%%%%%%%%
\newpage
\bibliographystyle{aasjournal}
\bibliography{ref}

%%%%%%%%%%%%%%%%%%%%%%%%%%%%%%%%%%%%%%%%%%%%%%%%%%
%
% APPENDIX
%
%%%%%%%%%%%%%%%%%%%%%%%%%%%%%%%%%%%%%%%%%%%%%%%%%%
\appendix
\section{Confidence level of Cross-matched Optical Counterparts}
\label{appendix:counterpart}

To determine the best-matched optical counterparts in the CAMIRA catalog, we develop the following framework to estimate the likelihood $P(N_\mathrm{mem}, \redshift_\mathrm{cl} | \nu)$ and choose the best-match counterpart to be the most likely. 
\begin{equation}
\label{eq:richness_likelihood}
    \begin{aligned}
        P(N_\mathrm{mem}, \redshift_\mathrm{cl} | \nu) &= \frac{1}{P(\nu)}\int P(N_\mathrm{mem}, \redshift_\mathrm{cl}, \nu, \mass)\,\dif\mass \\
        &=  \frac{1}{P(\nu)}\int P(N_\mathrm{mem}, \nu, | \mass, \redshift_\mathrm{cl})P(\mass, \redshift_\mathrm{cl})\,\dif\mass
    \end{aligned}
\end{equation}
To first-order approximation, we assume that the optical richness $N_{\mathrm{mem}}$ to be independent of the weak-lensing signal-to-noise ratio $\nu$ and write
\begin{equation}
    P\left(N_{\mathrm{mem}}, \nu| \mass, \redshift \right) = P\left(N_{\mathrm{mem}} | \mass, \redshift \right) P\left(\nu| \mass, \redshift \right).
\end{equation}
We make use of the richness--mass scaling relation from \citet{Murata2019} to evaluate $P\left(N_{\mathrm{mem}} | \mass, \redshift \right)$ as 
\begin{equation}
    P\left(N_{\mathrm{mem}} | \mass, \redshift \right) = \frac{1}{N_\mathrm{mem}\sqrt{2 \pi} \sigma_{\ln N \mid M, z}} \exp \left(-\frac{x^2(N_\mathrm{mem}, M, z)}{2 \sigma_{\ln N \mid M, z}^2}\right),
\end{equation}
%%%%%%%%%%%%%%%%%%%%%%%%%%%%%%%%%%%%%%%%%%%%%%%%%%
where $x(N, M, z)$ is given as
\begin{equation}
    x(N, M, z) \coloneqq \ln N_\mathrm{mem}-\left[A+B \ln \left(\frac{M}{M_{\text {pivot }}}\right)+B_z \ln \left(\frac{1+z}{1+z_{\text {pivot }}}\right)+C_z\left[\ln \left(\frac{1+z}{1+z_{\text {pivot }}}\right)\right]^2\right] .
\end{equation}
The redshift-dependent scatter $\sigma_{\ln N \mid M, z}$ is modeled as
\begin{equation}
\sigma_{\ln N \mid M, z} \coloneqq \sigma_0+q \ln \left(\frac{M}{M_{\mathrm{pivot}}}\right) +q_z \ln \left(\frac{1+z}{1+z_{\mathrm{pivot}}}\right)+p_z\left[\ln \left(\frac{1+z}{1+z_{\mathrm{pivot}}}\right)\right]^2.
\end{equation}
\citet{Murata2019} obtained constraints on the parameter $(A, B, B_z, C_z, \sigma_0, q, q_z, p_z)$ by fitting to the HSC-Y1 shear data assuming a pivot mass $M_{\mathrm{pivot}} = 3\times 10^{14}\,h^{-1}M_\odot$ and redshift $\redshift_{\mathrm{pivot}} = 0.6$. Here, we adopt the constraint derived under the \citet{Planck18} cosmology $(A, B, B_z, C_z, \sigma_0, q, q_z, p_z) = (3.36, 0.83, -0.20, 3.51, 0.19, -0.02, 0.23, 1.26)$. We note that this constraint is obtained for the optical clusters in the CAMIRA catalog with $N_\mathrm{mem} \geq 15$ and $0.1 \leq z_\mathrm{cl} \leq 1.0$. Therefore, we only cross-match to CAMIRA clusters within these ranges. For the other terms in Equation \eqref{eq:richness_likelihood}, $P\left(\nu| \mass, \redshift \right)$ is obtained following the treatment in Sec.~\ref{subsec:validation}
\begin{equation}
    P\left(\nu| \mass, \redshift \right) = \iint\dif\hat{\aperKappa} \dif\thetas
P( \nu | \hat{\aperKappa}, \thetas)
P(\hat{\aperKappa}, \thetas |\mass, \redshift, \vect{p}), 
\end{equation}
while $P(\mass, \redshift_\mathrm{cl})$ is derived from the halo mass function $\dif N / \dif \mass /\dif \redshift$ \citep{Bocquet16}. To be consistent with the richness--mass scaling relation, $P\left(\nu| \mass, \redshift \right)$ and $P(\mass, \redshift)$ are also derived under the \citet{Planck18} cosmology. 

\begin{figure}[tbh]
    \centering
    \epsscale{0.8}
    \plotone{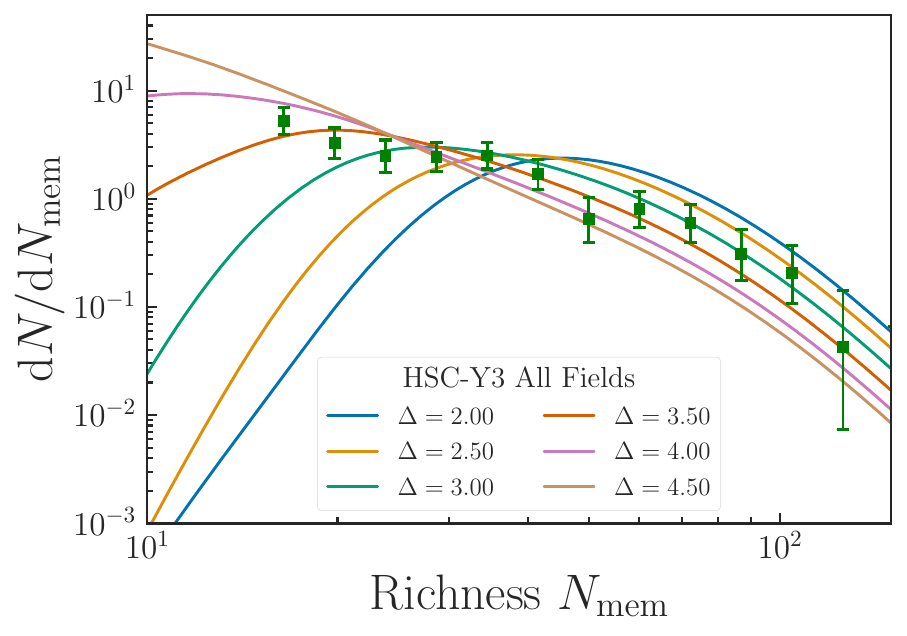}
    \caption{Same as \autoref{fig:richness_compare}, number counts of weak lensing peaks' optical counterparts as a function of their optical richness given the scaling relation $P(\nu | \hat{\aperKappa}, \thetas)$ derived with different allowed range of scattering $\Delta_s$. These results are derived with $P(\nu | \hat{\aperKappa}, \thetas)$ for HSC-Y3 after we unblind the cosmological analysis in \citet{PartII}. The green squares show the number counts of the closest CAMIRA counterpart for weak-lensing peaks in the entire HSC-Y3 footprint with a signal-to-noise ratio larger than $4.7$. \label{fig:richness_compare-unblind}}
\end{figure}

\end{document}